\documentclass[12pt,a4paper]{article}

\usepackage[a4paper,margin=2.5cm]{geometry}
\usepackage[round,authoryear]{natbib}
\usepackage{amsmath}
\usepackage{amssymb}
\usepackage{tikz}
\usetikzlibrary{arrows.meta,positioning}
\usepackage{array}
\usepackage{tabularx}
\usepackage{graphicx}
\usepackage{xcolor} 
\usepackage{placeins}
\usepackage{booktabs}
\usepackage[labelfont=bf]{caption}
\usepackage{longtable}
\usepackage{mathspec}
\setmathsfont(Digits,Latin,Greek)[
  Path = ./,
  UprightFont = SitkaLining.ttf,
  BoldFont = SitkaLining-Bold.ttf,
  ItalicFont = SitkaLining-Italic.ttf,
  BoldItalicFont = SitkaLining-BoldItalic.ttf
]{Sitka Lining}
\usepackage[hidelinks]{hyperref}

\usepackage{fancyhdr}

\fancypagestyle{firstpagepreprint}{
  \fancyhf{} % clear header and footer
  \fancyfoot[L]{Preprint, 3 May 2026}

}

\begin{document}

\thispagestyle{firstpagepreprint}

\begin{center}
{\fontsize{18}{22}\selectfont\bfseries Koopman Representations for Early Outbreak Warning and Minimal Counterfactual Intervention in Multi-Agent Epidemic Simulations
\par}
\vspace{1.5\baselineskip}
{\fontsize{14}{17}\selectfont\bfseries Florin Leon\par}
\vspace{1.0\baselineskip}
{\fontsize{11}{13}\selectfont Department of Computers\par}
{\fontsize{11}{13}\selectfont Faculty of Automatic Control and Computer Engineering\par}
{\fontsize{11}{13}\selectfont ``Gheorghe Asachi'' Technical University of Iași, Romania\par}
{\fontsize{11}{13}\selectfont Email: florin.leon@academic.tuiasi.ro\par}
\end{center}

\vspace{1.25\baselineskip}

\noindent{\fontsize{14}{17}\selectfont\bfseries Abstract}

\vspace{0.5\baselineskip}

This paper presents a Koopman-based framework for early outbreak detection and intervention selection in a multi-agent epidemic simulation. Agents exhibit mobility patterns, heterogeneous susceptibility, immunity-dependent viral load progression, and local transmission through co-location. The goal of the simulation is to study near-critical epidemic regimes in which small changes in exposure or timing can alter the final outcome. Aggregate daily observables from early trajectory windows are encoded into a low-dimensional Koopman latent space whose approximately linear evolution supports short-horizon forecasting and outbreak risk estimation. These representations are combined with a random forest classifier trained to predict whether the final attack rate exceeds a major outbreak threshold. Experiments near the system tipping points show strong early warning performance, with Koopman-derived features contributing to class separation. Counterfactual analysis further shows that minimal interventions, such as keeping a single selected agent at home for one day, can reduce attack rates and, often, shift the trajectory below the outbreak threshold.


\section{Introduction}

Many phenomena of scientific and social importance arise from complex systems composed of interacting entities. Epidemic spread, collective mobility, financial instability, ecological collapse, and information diffusion all exhibit macroscopic patterns that emerge from local interactions. Such systems are difficult to predict because their aggregate behavior is often nonlinear, path-dependent, and sensitive to small changes in initial conditions or local events. A central challenge is the identification of tipping points: regions of the system’s state space where a small perturbation can alter the long-term trajectory. Near such regions, the system may approach a phase transition, where outcomes that previously appeared stable become unstable and qualitatively different regimes are possible.

This paper studies tipping point detection and interventions in a multi-agent epidemic simulation. The epidemic setting provides a clear and socially meaningful example of a broader problem in complex systems. Disease transmission depends on individual mobility, heterogeneous biological response, local contact patterns, and temporal progression of infectiousness. These microscopic mechanisms interact to produce macroscopic outcomes such as extinction, limited spread, or large scale outbreak. In many parameter regimes, the final outcome is nearly predetermined: low transmissibility leads to rapid extinction, while high transmissibility leads to widespread infection. The most scientifically interesting regime lies between these extremes, where the system is close to a critical boundary and small changes in timing, exposure, or mobility can determine whether an outbreak occurs or not.

Traditional epidemic models offer valuable analytical structure, especially when populations can be represented through aggregate compartments such as susceptible, infected, and recovered groups. However, such models can obscure the role of individual heterogeneity, spatial contact structure, and contingent interaction histories. Agent-based models address these limitations by representing individuals, movement, local interactions, and disease progression directly. However, their flexibility comes with a cost: the resulting dynamics are high-dimensional, nonlinear, and often difficult to summarize. Simulation can reveal what happens in a given run, yet it does not automatically provide an early warning representation of where the system is heading or which minimal intervention could change its outcome.

Koopman operator methods provide one route toward this problem. The Koopman perspective represents nonlinear dynamics through the evolution of observables in a lifted space. Although the underlying system may be nonlinear, the evolution of suitable observables can often be approximated by linear operators in a learned latent representation. This property makes Koopman-based models attractive for complex dynamical systems, where direct mechanistic prediction may be difficult, and pure black box forecasting may provide limited insights. In the present setting, the goal is to learn a compact representation of early epidemic trajectories that preserves information relevant to long-term outcomes.

The proposed study combines three components. First, a multi-agent epidemic simulator generates trajectories from individual mobility, viral-load progression, susceptibility, and local transmission. Second, a Koopman-inspired model learns latent representations of early aggregate epidemic dynamics. These representations support short-horizon forecasting and provide features for outbreak classification. Third, counterfactual intervention experiments test whether minimal mobility restrictions can alter the final outcome when the system lies near a tipping point. The intervention considered here is deliberately simple: a selected agent is kept at home for a selected day, after which the simulation proceeds normally. This design isolates the effect of a small, localized perturbation on the global trajectory.

The central hypothesis is that systems near critical epidemic transitions contain exploitable structure. If early dynamics indicate proximity to an outbreak boundary, then targeted intervention may prevent a transition to widespread infection. Conversely, when the system is far from the boundary, the same intervention may have little effect because the trajectory is already strongly attracted toward extinction or outbreak. The value of the approach therefore lies in the joint use of early prediction and counterfactual simulation: Koopman representations help identify likely outcomes from short observation windows, while intervention search tests whether those outcomes are modifiable.

The broader contribution of the paper is a computational framework for studying phase-transition-like behavior in multi-agent systems. Epidemic spread is used as the motivating domain, yet the conceptual framework extends to other systems where local interactions generate global regimes and where timely, minimal interventions may change system evolution.

The remainder of the article is organized as follows. Section 2 reviews related work on epidemic cascades, tipping point analysis, Koopman operator learning, and counterfactual intervention in multi-agent systems. Section 3 introduces the simulation model, including the agent population, mobility and contact formation, disease progression, transmission mechanism, aggregate observables, Koopman representation, early warning classifier, and intervention model. Section 4 presents the experimental design and results that focus on boundary-focused simulations, latent Koopman features, outbreak prediction, and quarantine counterfactuals. Section 5 concludes the paper and outlines some future research directions.

The implementations of the model and the experiments are available at: {\color{blue}https://github.com/ florinleon/KoopmanEpidemics}.

\section{Related Work}

Research on cascades in networks provides a natural starting point for epidemic outbreak analysis. \citet{watts2002global} develops a threshold model in which agents switch state when the fraction of active neighbors exceeds an individual threshold. The model shows that large cascades can arise from small initial shocks, but only when the interaction network lies inside a vulnerable structural regime. Its main contribution is to connect microscopic exposure rules, threshold heterogeneity, and network degree distribution to macroscopic cascade size. The work is especially relevant for simulations near critical boundaries because it explains why similar seeds can produce very different outcomes, and why cascade susceptibility is a property of the population structure rather than of the initiating event alone.

The analogy between epidemic waves and finite-size phase transitions is developed by \citet{das2023finite}. The paper studies COVID-19 infection and fatality trajectories and argues that many regional curves collapse under a compact finite-size scaling form. This perspective treats epidemic growth as a process with an early exponential regime, a later constrained regime, and a saturation level set by system size, interventions, immunity, and behavioral restrictions. The comparison with coarsening dynamics in finite statistical-physics systems gives a useful language for near-threshold epidemic behavior. It also suggests that attack rate outcomes should be interpreted not only through reproduction intensity, but also through saturation, finite-population effects, and constraints that redirect trajectories before they reach their final size.

Node-level intervention in epidemic networks is addressed by \citet{farahi2024critical}, who study critical node detection in temporal social networks. They propose temporal supracycle ratio, temporal semi-local integration, and temporal semi-local centrality, and evaluate these measures under SIR (Susceptible, Infected, Recovered) spreading, isolation, and robustness experiments. The work emphasizes that influential nodes cannot be reliably characterized only by static degree when contacts change over time. A node can be epidemiologically important because it connects temporal paths, appears at critical moments, or bridges local neighborhoods during the infectious period. This result is important for intervention design because it changes attention from static ranking of individuals to agent importance over time.

A related control-theoretic view appears in \citet{zhu2019disturbance}, who study disturbance rejection for discrete-time multi-agent systems with nonlinear couplings, random switching topologies, and exogenous disturbances. Their analysis uses Lyapunov methods to derive sufficient conditions for bounded disturbance response, and the experiments use pinning control to influence selected agents. Although the application domain is multi-robot and consensus control rather than epidemiology, the conceptual connection is that collective behavior can sometimes be steered through carefully selected local interventions. The paper illustrates why limited control of key agents may be more practical than population-wide control, especially when interactions are nonlinear and network topology changes over time.

The analysis of tipping pathways in high-dimensional agent-based models is advanced by \citet{helfmann2021statistical}. They combine diffusion maps with transition path theory to study noise-induced transitions between relevant regions of agent-state space. Diffusion maps recover low-dimensional collective coordinates from simulation data, while transition path theory quantifies pathway probabilities, reactive fluxes, transition rates, and mean transition durations. Their experiments show that tipping points in agent systems may involve several competing routes rather than one deterministic threshold crossing. This framework is significant because it treats tipping as an ensemble of possible paths through a reduced state space.

Task-oriented surrogate modeling for tipping points in agent-based systems is developed by \citet{fabiani2024task}. The paper combines manifold learning, neural networks, Gaussian processes, and equation-free multiscale analysis to construct reduced-order models for detailed agent-based simulators. Its experiments include a stochastic financial-market and a stochastic epidemic on an Erd\H{o}s--R\'{e}nyi network. A central finding is that, near the tipping point, the emergent dynamics of both systems can be represented by an effective one-dimensional stochastic differential equation. The work is important because it argues that the modeling target should determine the surrogate: for tipping detection, a compact task-specific representation may be more useful than a full microscopic emulator.

Epidemiological tipping points in adaptive networks is studied by \citet{evangelou2024tipping}. The paper uses reduced stochastic differential equations assisted by machine learning to capture coarse epidemic dynamics in an adaptive network. A ResNet-inspired architecture is used to learn drift and diffusion terms, and diffusion maps help identify coarse observables when hand-designed variables are insufficient. The resulting model supports bifurcation analysis and rare-event estimation for transitions toward large-amplitude collective behavior. This work connects epidemic dynamics, dimensionality reduction, and stochastic coarse modeling, and also reinforces the idea that epidemic tipping points can be described by a low-dimensional process embedded in a much higher-dimensional network state.

Early warning indicators based on reconstructed network geometry are proposed by \citet{peng2020identification}. Their method transforms time series into complex networks through quasi-isometric mapping, then uses changes in the resulting network structure to identify and predict bifurcation tipping points. The paper belongs to a broader class of methods that search for pre-transition signatures in transformed representations rather than in raw trajectories. Its main contribution is to show that geometric and topological properties of reconstructed dynamics may become informative before a visible macroscopic transition occurs. This is relevant for nonlinear systems in which conventional scalar indicators are weak, noisy, or delayed relative to the underlying process toward instability.

Machine learning prediction for tipping transitions is explored by \citet{patel2023machine}. The work considers nonstationary dynamical systems in which parameters drift and a model must infer the approach to a tipping point from pre-transition data. It also studies extrapolation beyond the transition, which is difficult because the post-tipping regime may differ qualitatively from the training regime. The paper highlights a key challenge for data-driven early warning, i.e., observations before a transition can be ambiguous, yet they must contain enough information to infer both the imminent threshold and the likely subsequent state. Its contribution is to frame tipping prediction as a representation and extrapolation problem under nonstationarity.

Efficient extrapolation of nonstationary complex dynamics is further developed by \citet{koglmayr2024extrapolating}. The paper proposes a machine learning framework for forecasting tipping points and simulating dynamics when the governing equations are unavailable or incomplete. It focuses on the difficult case where the system evolves outside the training distribution as it approaches a critical transition. The method demonstrates that reduced learned representations can support both tipping point extrapolation and forward simulation beyond the immediate data window. This line of work is important for early warning applications because decisions must often be made before the final regime is observable.

The operator-theoretic foundation for Koopman-based modeling is surveyed by \citet{budisic2012applied}. The Koopman perspective replaces nonlinear state evolution with linear evolution of observable functions. Although the state dynamics may be nonlinear, the induced action on observables is linear in an infinite-dimensional function space. This view provides a principled way to study nonlinear systems through spectra, modes, and coherent structures. Its appeal for data-rich complex systems lies in the possibility of finding finite sets of observables that behave approximately linearly over useful horizons. The survey clarifies both the mathematical basis of the Koopman operator and its role as a bridge between nonlinear dynamics, spectral analysis, and data-driven reduction.

A practical approximation of Koopman dynamics is given by \citet{williams2015data} through extended dynamic mode decomposition. Extended Dynamic Mode Decomposition (EDMD) lifts state snapshots into a dictionary of observables and fits a finite-dimensional linear operator that advances those observables. The method generalizes dynamic mode decomposition and connects data-driven regression to Koopman and Perron--Frobenius operator approximation. Its importance is methodological, as nonlinear dynamics can be forecast by linear evolution after an appropriate feature map. The paper also clarifies how the choice of dictionary controls approximation quality. This makes EDMD a central reference for later methods that replace hand-designed dictionaries with learned latent coordinates or task-specific observable maps.

Deep Koopman learning is studied by \citet{lusch2018deep}, who use neural networks to discover coordinates in which nonlinear dynamics evolve approximately linearly. The architecture learns an encoder, a linear latent evolution model, and a decoder, and it can handle systems with continuous spectra through auxiliary networks. This work is central because it moves Koopman modeling from fixed observable dictionaries toward learned representations. It shows that a compact latent space can preserve predictive dynamical structure when the coordinates are chosen from data. The paper also presents a useful criterion for representation learning in dynamical systems: a latent space should not merely reconstruct states, but should also support simple and accurate temporal evolution.

The connection between Koopman representations and control is analyzed by \citet{brunton2016koopman}. The paper studies finite-dimensional Koopman-invariant subspaces and discusses when nonlinear systems can be represented by linear dynamics in the lifted observable space. It also explains the limitations of exact finite-dimensional closure and the practical value of approximate invariant observables for prediction and control. This contribution links representation quality to intervention design: a useful lifted model should support not only forecasting but also reasoning about how actions alter future trajectories. The work therefore connects Koopman theory with model reduction, nonlinear control, and data-driven decision support.

Interpretability of learned latent spaces is examined by \citet{leeb2022exploring}. The authors propose interventional assays for autoencoder representations, where latent coordinates are perturbed and the resulting responses are measured. Their framework separates informative structure from noise, analyzes relationships among latent variables, and evaluates how the learned representation encodes underlying factors of variation. The paper shoes that a low-dimensional latent space is not automatically interpretable or causally meaningful. Good downstream accuracy can coexist with entangled, fragile, or geometrically misleading coordinates. Interventional latent analysis therefore provides a complementary diagnostic for representation learning, especially when latent variables are later used for forecasting, classification, or intervention.

Counterfactual agent importance in multi-agent systems is studied by \citet{chen2024understanding}. Their method estimates individual importance by measuring how much global reward changes when selected agents' actions are randomized. Because exhaustive counterfactual perturbation is infeasible in large multi-agent systems, the method learns a masking procedure that identifies agents whose actions can be randomized while preserving reward. Important agents are those whose perturbation causes large outcome changes. Although the setting is multi-agent reinforcement learning, the principle generalizes to simulations where global outcomes depend on local agent behavior. The work provides a direct methodological link between counterfactual reasoning, agent-level attribution, and targeted intervention under interaction effects.

These works establish a methodological chain where network cascade models explain why local exposure rules can produce abrupt macroscopic transitions, epidemic and agent-based modeling studies show how such transitions can be analyzed through low-dimensional collective variables. Tipping point literature develops early warning and extrapolation methods for nonlinear systems, Koopman theory supplies a linear latent-dynamics framework for forecasting, and counterfactual multi-agent methods connect prediction to agent-level intervention and attribution.

\section{Model Description}

\subsection{Scope and Formulation}

The proposed model is a multi-agent simulation of epidemic spread designed to study early outbreak detection and minimal intervention near tipping points. The model represents an artificial population whose members move through a shared spatial environment, interact through local co-location, and undergo individual disease progression after infection. Its purpose is to provide a controlled computational system in which nonlinear epidemic trajectories can be generated, predicted from early observations, and modified through counterfactual interventions.

The main object of study is the transition between contained and outbreak regimes. In low-transmission regimes, an initially infected agent produces few secondary infections and the disease disappears rapidly. In high-transmission regimes, the infection spreads through a large fraction of the population and intervention after the early phase has limited effect. The most relevant regime lies near the boundary between these outcomes. In this region, small changes in local mobility or exposure can alter the final epidemic state. The simulation is therefore used as a testbed for studying phase-transition-like behavior in a complex adaptive system.

Let
$\mathcal{A}=\{1,\ldots,N\}$
denote the set of agents. Each agent \(i \in \mathcal{A}\) has fixed attributes assigned at initialization, including a home location, a daily movement routine, a susceptibility value, and an immunity category. Each agent also has dynamic variables, including location, epidemiological state, infection age, and viral load. The spatial environment consists of a finite grid
$\mathcal{G}=\{0,\ldots,G-1\}\times \{0,\ldots,G-1\}$
and a set of off-grid homes
$\mathcal{H}=\{1,\ldots,H\}$.

At each simulation step, an agent occupies either a grid cell \(g \in \mathcal{G}\) or a home \(h \in \mathcal{H}\). Co-location creates the opportunity for transmission. The general model permits transmission in both grid cells and homes. The experimental results evaluate transmission only in grid cells, which simplifies the contact process and makes the resulting tipping behavior easier to isolate. The same prediction and intervention framework can be applied when home transmission is enabled, although the epidemic dynamics would then become more aggressive because household exposure would create an additional contact channel.

Time is discrete. A simulation day consists of \(L=L_D+L_N\) steps, where \(L_D\) denotes the number of daytime movement steps and \(L_N\) denotes the number of nighttime home steps. The simulator state at step \(t\) can be written as:

\begin{equation}
X_t =
\left(
x_1(t),\ldots,x_N(t)
\right),
\end{equation}

\noindent where \(x_i(t)\) contains the full internal state of agent \(i\). For a fixed parameter vector \(\theta\), random seed \(\omega\), and intervention input \(u_t\), the system evolves according to a discrete transition rule:

\begin{equation}
X_{t+1}=F_{\theta}(X_t,u_t;\omega).
\end{equation}

The intervention input \(u_t\) is inactive in baseline simulations. In counterfactual simulations, it modifies the mobility of selected agents at selected times. Because each run uses fixed initialization and a fixed random seed, the simulator defines a reproducible trajectory for each parameter setting and intervention schedule. This property is essential for counterfactual analysis: a baseline run and an intervention run can be compared under identical initial conditions and identical stochastic choices, so outcome differences can be attributed to the intervention.

The simulator records aggregate daily observables rather than the full microscopic state for the prediction task. Let
$Y_d = \phi(X_{dL})$
denote the vector of end-of-day observables on day \(d\). These observables include the number of susceptible, infected, recovered, and dead agents, as well as daily incident infections and mobility-relevant infected counts. The learning problem is formulated on short histories of these aggregate observables. Given an early observation window
${Y}_{d-k+1:d}=(Y_{d-k+1},\ldots,Y_d)$,
the predictive component estimates whether the trajectory will end in a contained regime or a major outbreak. The final outcome is defined by the attack rate:

\begin{equation}
\rho = \frac{N-S_T}{N},
\end{equation}

\noindent where \(S_T\) is the number of susceptible agents at the end of the simulation horizon. A major outbreak is assigned when \(\rho\) exceeds a chosen threshold \(\rho_c\). In the reported experiments, \(\rho_c=0.3\).

This formulation separates three levels of the method. The first level is the microscopic agent-based simulator, which produces epidemic trajectories from movement, contact, viral load, and susceptibility. The second level is the early warning model, which uses short aggregate histories to infer the likely final regime. The third level is the counterfactual intervention procedure, which tests whether a small mobility restriction can move a near-critical trajectory from the outbreak regime into the contained regime. The scientific focus is therefore not only outbreak prediction, but the identification of trajectories whose outcomes remain modifiable because the system lies close to a tipping boundary.

\subsubsection*{Specific Details}

In the model description and experimental sections, we include dedicated subsections that specify all parameters used in the experiments, together with their default values. Although these implementation details could have been placed in an appendix, we present them close to the corresponding conceptual descriptions to provide a more complete and, perhaps, more intuitive view of each part of the model. Readers who are primarily interested in the conceptual overview may skip these subsections.

\begin{longtable}{@{}p{0.15\textwidth} p{0.75\textwidth}@{}}
\textbf{Symbol} & \textbf{Meaning} \\
\hline
\(N\) & Number of agents \\
\(\mathcal{A}\) & Set of agents, \(\{1,\ldots,N\}\) \\
\(G\) & Width and height of the square spatial grid \\
\(\mathcal{G}\) & Set of grid cells \\
\(H\) & Number of home locations \\
\(\mathcal{H}\) & Set of homes \\
\(L_D\) & Number of daytime steps per day \\
\(L_N\) & Number of nighttime steps per day \\
\(L\) & Total number of simulation steps per day, \(L=L_D+L_N\) \\
\(t\) & Simulation step index \\
\(d\) & Simulation day index \\
\(X_t\) & Full microscopic simulator state at step \(t\) \\
\(x_i(t)\) & Full state of agent \(i\) at step \(t\) \\
\(Y_d\) & Aggregate end-of-day observable vector \\
\(\theta\) & Vector of simulator parameters \\
\(\omega\) & Random seed or stochastic initialization identifier \\
\(u_t\) & Intervention input at step \(t\) \\
\(F_\theta\) & Simulator transition rule \\
\(\phi\) & Observation map from microscopic states to aggregate daily observables \\
\(S_T\) & Final number of susceptible agents \\
\(\rho\) & Final attack rate \\
\(\rho_c\) & Major outbreak threshold \\
\end{longtable}

The baseline specification uses a \(50 \times 50\) spatial grid, \(H=200\) home locations, and \(N=500\) agents. Each day contains \(L_D=10\) daytime movement steps followed by \(L_N=10\) nighttime home steps, so \(L=20\). The model has 60 days as the default experimental horizon, while the implementation also supports, e.g., evaluation horizons of 365 days. The experiments use 60-day runs to focus on early outbreak detection and to avoid unnecessary computation after the epidemic outcome has effectively been determined.

The final attack rate is computed as
$\rho={(N-S_T)} / {N}$.
A run is labeled as a major outbreak when
$\rho \geq 0.3$.
This threshold is used as the supervised label for the outbreak-classification task and as the criterion for evaluating whether an intervention prevented an outbreak.

\subsection{Agent Population and the Environment}

The simulator represents an epidemic process in a finite artificial population situated in a spatial interaction environment. The population consists of \(N\) agents,
$\mathcal{A}=\{1,\ldots,N\}$,
and the environment consists of two types of locations. The first is a square grid \(\mathcal{G}\), which represents the shared daytime interaction space. The second is a set of off-grid home locations \(\mathcal{H}\), which represents private nighttime residence locations. At any simulation step, each agent occupies exactly one location in
$\mathcal{L}=\mathcal{G}\cup \mathcal{H}$.
The location of agent \(i\) at step \(t\) is denoted by \(\ell_i(t)\in \mathcal{L}\).

The separation between grid locations and homes allows the model to distinguish between public and residence-based interactions. During the active part of the day, agents move through grid cells according to individual routines. During the nighttime, agents return to their assigned homes. This design creates repeated but heterogeneous contact patterns: agents do not mix uniformly, and their exposure histories depend on the overlap between individual routines, disease progression, and timing.

Each location may contain any number of agents. Thus, the model does not impose a physical capacity limit on grid cells or homes. This assumption makes co-location the only condition required for possible interaction. If several agents occupy the same location at the same simulation step, the location defines a temporary contact group. For a location \(q\in \mathcal{L}\), the group present at time \(t\) is:

\begin{equation}
C_q(t)={i\in\mathcal{A}:\ell_i(t)=q}.
\end{equation}

Transmission can then be evaluated within such groups according to the disease-transmission rule described later.

At initialization, each agent receives a fixed home assignment,
$h_i\in\mathcal{H}$,
and a daily grid routine,
$R_i=(r_{i,0},r_{i,1},\ldots,r_{i,K-1}), r_{i,j}\in\mathcal{G}$.
The routine defines the grid cells visited by the agent during the daytime. The routine is assigned once and remains fixed during a baseline simulation, although the starting point within the routine changes across days through a phase parameter described in the mobility subsection. The fixed routine assumption gives each agent a persistent spatial signature. This is important because epidemic spread in the model is determined not only by the number of agents and the transmission threshold, but also by the repeated overlaps between specific agents’ movement patterns.

Each agent also receives biological attributes that affect its response to infection. The susceptibility parameter, denoted by \(s_i\), scales the exposure required for agent \(i\) to become infected. The immunity category, denoted by \(m_i\), determines the viral-load curve followed by the agent after infection. These attributes create heterogeneity at the individual level. Two agents with the same contact history may therefore have different infection outcomes, and two infected agents may produce different future risks because their viral-load trajectories may differ.

The epidemiological state of each agent is initialized as susceptible except for a single randomly selected index \(i_0\), which is marked infected at the beginning of the simulation:

\begin{equation}
z_i(0)=
\begin{cases}
I, & i=i_0,\\
S, & i\neq i_0,
\end{cases}
\end{equation}

\noindent where \(z_i(t)\in\{S,I,R,D\}\) denotes the disease state of agent \(i\) at time \(t\). The states \(S\), \(I\), \(R\), and \(D\) denote susceptible, infected, recovered, and dead, respectively. The initial infected agent begins the epidemic process, and all subsequent infections arise through contact dynamics.

Initialization is controlled by a single random seed. The seed determines home assignments, movement routines, phase values, susceptibility values, immunity categories, and the initially infected agent. Once these quantities have been generated, the baseline trajectory is deterministic under the fixed model parameters. This design supports counterfactual comparison: an intervention run can reuse the same initialized population and differ only in the imposed intervention. Consequently, differences between baseline and intervention trajectories can be interpreted as effects of the intervention within the simulated system.

The population and environment specification is intentionally abstract. The agents do not represent named individuals, specific demographic groups, or empirical households. The grid does not represent a calibrated physical geography. The purpose of this abstraction is to create a controlled complex system in which local co-location, heterogeneous susceptibility, individual disease progression, and mobility timing jointly determine the macroscopic epidemic outcome. This level of abstraction is appropriate for the central objective of the paper: identifying early signals of near-critical epidemic trajectories and testing whether small counterfactual changes can redirect them.

\subsubsection*{Specific Details}

The following quantities define the baseline population and spatial environment used in the reference configuration.

\begin{longtable}{@{}p{0.2\textwidth} p{0.1\textwidth} p{0.17\textwidth} p{0.53\textwidth}@{}}
\textbf{Quantity} & \textbf{Symbol} & \textbf{Baseline value} & \textbf{Description} \\
\hline
Number of agents & \(N\) & 500 & Size of the simulated population \\
Grid width & \(G\) & 50 & Number of grid cells per spatial dimension \\
Grid cells & \(\mathcal{G}\) &  &  \\
Home locations & \(H\) & 200 & Off-grid residential locations \\
Routine length & \(K\) & 10 & Number of distinct grid cells in each \newline agent’s daily routine \\
Initial infected \newline agents &  & 1 & Number of infected agents at simulation \newline start \\
\end{longtable}

The grid is defined as: 
$\mathcal{G}=\{0,\ldots,G-1\}\times \{0,\ldots,G-1\}$.
The home set is defined as:
$\mathcal{H}=\{1,\ldots,H\}$.
Each agent \(i\) is assigned one home \(h_i\in\mathcal{H}\). Home assignment is random and does not enforce equal household sizes. Therefore, the realized distribution of home occupancy depends on the random seed. Since homes are modeled as locations with unlimited capacity, several agents may share the same home.

Each agent is also assigned a routine \(R_i\) consisting of \(K\) distinct grid cells. The cells are selected randomly at initialization. The distinctness condition prevents an agent from revisiting the same grid cell within a single daily routine, although the same cell may appear in the routines of many different agents. These overlaps are one of the mechanisms through which persistent contact structure arises.

A compact initialization procedure is as follows:

\begin{equation*}
\begin{aligned}
h_i &\sim \mathrm{Uniform}(\mathcal{H}),\\
R_i &\sim \mathrm{SampleDistinct}(\mathcal{G},K),\\
s_i &\sim \mathcal{D}_s,\\
m_i &\sim \mathcal{D}_m,\\
p_i &\sim \mathrm{Uniform}{(1,\ldots,K-1)},
\end{aligned}
\end{equation*}

\noindent for each \(i\in\mathcal{A}\). Here \(p_i\) is the phase parameter used in the daily mobility rule, \(\mathcal{D}_s\) is the susceptibility distribution, and \(\mathcal{D}_m\) is the immunity-category distribution.

The reference configuraton uses:

\begin{equation*}
s_i \sim \mathrm{Uniform}(0.5,1.5).
\end{equation*}

The experimental study uses narrower susceptibility intervals when the objective is to sample near the transition between contained and outbreak regimes. These narrow intervals are an experimental design choice used to concentrate simulations near tipping points, where interventions are most informative.

\subsection{Mobility and Contact Formation}

Agent mobility determines the contact structure of the epidemic process. The model uses daily routines rather than random movement at each step. This choice creates persistent yet non-identical interaction patterns: agents repeatedly visit the same personal set of grid cells, but their position within the routine changes between days. As a result, contacts arise from the temporal overlap between routines, phase shifts, and disease-dependent movement restrictions.

Time is discrete. Let \(d\in\{0,1,\ldots,T-1\}\) denote the day index and let \(\tau\in\{0,\ldots,L-1\}\) denote the within-day step index. The corresponding global step is:

\begin{equation}
t=d \cdot L+\tau,
\end{equation}

\noindent where:

\begin{equation}
L=L_D+L_N.
\end{equation}

The first \(L_D\) steps of each day are daytime movement steps, and the remaining \(L_N\) steps are nighttime home steps. During daytime, agents may visit grid locations. During nighttime, agents are assigned to their homes.

Each agent \(i\) has a fixed routine:

\begin{equation}
R_i=(r_{i,0},r_{i,1},\ldots,r_{i,K-1}),
\end{equation}

\noindent where each \(r_{i,j}\in\mathcal{G}\). The routine contains the grid locations available to the agent during the daytime part of each day. A phase parameter \(p_i\) controls where the agent begins within this routine on a given day. The daily offset is:

\begin{equation}
o_i(d)=(d \cdot p_i)\bmod K.
\end{equation}

For a daytime step \(\tau<L_D\), the nominal routine index is:

\begin{equation}
j_i(d,\tau)=(o_i(d)+\tau)\bmod K,
\end{equation}

\noindent and the nominal daytime location is:

\begin{equation}
\ell_i^{0}(d,\tau)=r_{i,j_i(d,\tau)}.
\end{equation}

Thus, agents do not always start the day at the same position in their routine. The phase-shift mechanism prevents the contact network from repeating exactly every day, while preserving agent-specific spatial regularity. This gives the simulation a simple form of temporal heterogeneity: the same pair of agents may meet on some days and not on others, depending on the alignment of their routines and phases.

During nighttime steps, the nominal location of agent \(i\) is its assigned home:

\begin{equation}
\ell_i^{0}(d,\tau)=h_i,\qquad L_D\leq \tau<L.
\end{equation}

The full location rule also depends on epidemiological and intervention states. Let \(b_i(t)\in\{0,1\}\) denote whether agent \(i\) is behaviorally restricted by the disease process at step \(t\), and let \(q_i(t)\in\{0,1\}\) denote whether the agent is restricted by an imposed intervention. If either variable is active, the agent remains at home:

\begin{equation}
\ell_i(t)=h_i
\quad \text{if} \quad
b_i(t)=1 \ \text{or}\ q_i(t)=1.
\end{equation}

Otherwise, the agent follows the nominal location rule:

\begin{equation}
\ell_i(t)=\ell_i^{0}(d,\tau).
\end{equation}

Dead agents are removed from movement and transmission processes. They therefore do not contribute to subsequent contact groups. Recovered agents resume ordinary mobility unless an intervention restricts them, although they cannot be reinfected.

Contact formation is defined by co-location. At each step \(t\), every location \(q\) induces a contact group:

\begin{equation}
C_q(t)={i\in\mathcal{A}: \ell_i(t)=q}.
\end{equation}

If \(|C_q(t)|\geq 2\), agents in that group are considered co-located for that step. The model does not explicitly represent distance within a cell, duration within a step, mask use, ventilation, or pairwise social relationships. These factors are abstracted into the transmission rule and the model parameters. The assumption is that transmission opportunities arise from simultaneous presence at the same location.

The general simulator can treat both grid cells and homes as contact locations, but the current experiments restrict transmission evaluation to daytime grid cells. Nighttime home assignment still affects the agent state because symptomatic or intervened agents remain home and therefore do not participate in grid contacts. However, in the reported experimental configuration, home co-location itself does not generate new infections. This simplifies the interpretation of intervention effects: keeping an agent at home removes that agent from the daytime contact process without adding a compensating household transmission channel.

The mobility model is intentionally simple, but it captures several features that are important for studying tipping behavior. Contact opportunities are given by individual routines rather than fully homogeneous mixing. Contact timing varies between days through phase shifts. Disease progression feeds back into mobility by removing highly symptomatic agents from the grid. These properties make the epidemic trajectory sensitive to who is infectious, where they are located, and when their contacts occur.

\subsubsection*{Specific Details}

The baseline implementation uses the following daily timing structure.

\begin{longtable}{@{}p{0.25\textwidth} p{0.1\textwidth} p{0.17\textwidth} p{0.45\textwidth}@{}}
\textbf{Quantity} & \textbf{Symbol} & \textbf{Baseline value} & \textbf{Description} \\
\hline
Routine length & \(K\) & 10 & Number of grid cells in each agent’s routine \\
Daytime steps per day & \(L_D\) & 10 & Number of routine visits per day \\
Nighttime steps per day & \(L_N\) & 10 & Number of home-residence steps per day \\
Total steps per day & \(L\) & 20 & \(L_D+L_N\) \\
Phase values & \(p_i\) & \(\{1,\ldots,9\}\) & Daily routine shift values \\
\end{longtable}

In the reference configuration, \(K=L_D=10\). Therefore, an unrestricted agent visits every location in its routine once during each daytime period. The phase parameter changes the order of those visits across days. Since \(p_i\in\{1,\ldots,K-1\}\), the offset changes from one day to the next rather than remaining fixed. For \(K=10\), different phase values generate different cyclic patterns through the routine.

The disease-dependent mobility restriction is determined by viral load. Let \(v_i(t)\) denote the viral load of infected agent \(i\) at step \(t\), expressed on the scale used by the viral-load curves. The baseline behavioral rule is:

\begin{equation}
b_i(t)=
\begin{cases}
1, & z_i(t)=I \ \text{and}\ v_i(t)>50,\\
0, & \text{otherwise},
\end{cases}
\end{equation}

\noindent except that dead agents are handled separately and removed from the simulation dynamics. Agents with viral load below this homebound threshold follow their ordinary daytime routines. Agents above the threshold remain at home.

\subsection{Disease States and Viral Load Progression}

Disease progression is represented at the agent level. Each agent has an epidemiological state
$z_i(t)\in\{S,I,R,D\}$,
where \(S\) denotes susceptible, \(I\) infected, \(R\) recovered, and \(D\) dead. Susceptible agents can become infected through contact with infectious agents. Infected agents follow an immunity-dependent viral load trajectory. Recovered agents are considered immune to reinfection. Dead agents are removed from the simulation.

The disease process is driven by the viral load of infected agents. Let
$v_i(t)\geq 0$
denote the viral load of agent \(i\) at simulation step \(t\). For susceptible and recovered agents, viral load is treated as zero for transmission and behavioral purposes. For infected agents, viral load evolves according to a predefined curve associated with the agent’s immunity category. The uploaded specification defines four immunity categories, each mapped to a separate viral-load curve loaded from external data files.

Each agent receives an immunity category \(m_i\) during initialization. This category determines the function
$\gamma_{m_i}(a)$,
where \(a\) is infection age, measured in simulation steps since the agent became infected. If agent \(i\) is infected at step \(t_i^{\mathrm{inf}}\), then its infection age at step \(t\geq t_i^{\mathrm{inf}}\) is:

\begin{equation}
a_i(t)=t-t_i^{\mathrm{inf}}.
\end{equation}

The viral load is then given by:

\begin{equation}
v_i(t)=\gamma_{m_i}(a_i(t)).
\end{equation}

\begin{figure}[htbp]
    \centering
    \includegraphics[width=0.8\textwidth]{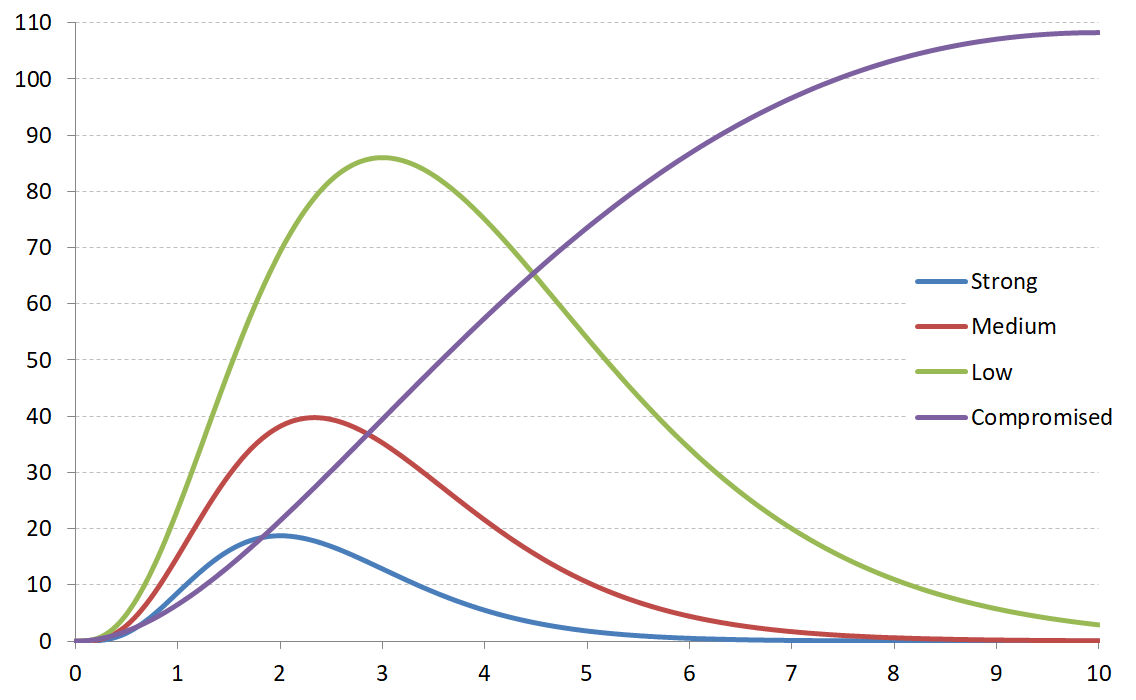}
    \caption{Viral load curves}
    \label{fig1}
\end{figure}

The curves encode different biological responses to infection. As shown in Figure~\ref{fig1}, strong immunity corresponds to a lower or shorter viral load trajectory, while weaker immunity can produce higher or more prolonged viral loads. The model therefore avoids assigning identical infectiousness profiles to all infected agents. This heterogeneity is important because the epidemic trajectory can depend strongly on which agents become infected early and how their infectious periods align with their mobility routines.

Viral load affects both transmission potential and behavior. First, higher viral load increases the ability of an infected agent to transmit disease during co-location. The transmission rule is described in the next subsection. Second, viral load determines whether the agent continues ordinary movement, becomes symptomatic, becomes homebound, recovers, or dies.

The behavioral and state-transition logic is defined by thresholds. If an infected agent’s viral load is below the symptom threshold, the agent behaves normally. If the viral load is above the symptom threshold but below the homebound threshold, the agent is symptomatic but continues to follow its daytime routine. If the viral load exceeds the homebound threshold, the agent remains at home and no longer participates in ordinary daytime movement. If the viral load exceeds the death threshold, the agent enters state \(D\) and is removed from further movement and transmission processes. These rules couple within-host progression to population-level spread: an agent with high viral load may become highly infectious, but sufficiently high disease severity also reduces that agent’s mobility.

Recovery occurs when the viral load remains below the recovery threshold through the end of the day. More precisely, after an infected agent’s viral load falls below the recovery threshold, the agent must remain below that threshold until the end of the current nighttime period. The agent is then marked recovered. This rule avoids instantaneous recovery in the middle of a day and ensures that state transitions occur consistently with the daily structure of the simulation. Once recovered, the agent cannot be reinfected.

\subsubsection*{Specific Details}

The immunity category is assigned at initialization as
$m_i\in{\text{strong},\text{medium},\text{low},\text{compromised}}$.
The reference immunity probabilities are:

\begin{equation*}
P(m_i=\text{strong})=0.35,
\end{equation*}

\begin{equation*}
P(m_i=\text{medium})=0.45,
\end{equation*}

\begin{equation*}
P(m_i=\text{low})=0.15,
\end{equation*}

\begin{equation*}
P(m_i=\text{compromised})=0.05.
\end{equation*}

Let \(v_i(t)\) denote viral load. The threshold rules are:

\begin{longtable}{@{}p{0.5\textwidth} p{0.5\textwidth}@{}}
\textbf{Viral-load range} & \textbf{Agent behavior or state transition} \\
\hline
\(v_i(t)<10\) & Infected but behaviorally normal \\
\(10\leq v_i(t)\leq 50\) & Symptomatic but still mobile \\
\(v_i(t)>50\) & Homebound; does not follow daytime routine \\
\(v_i(t)>100\) & Dead; removed from simulation \\
\(v_i(t)<1\) through the end of the current day & Recovered at the end of the nighttime period \\
\end{longtable}

The death transition is: $z_i(t^+)=D$ if $z_i(t)=I$ and $v_i(t)>100$.
The recovery transition is applied at the end of the nighttime segment. If
$v_i(t')<1$
for all relevant steps $t'$ from the first below-threshold step through the end of the current day, then
$z_i(t_{\mathrm{end}}^+)=R$.

The viral load interpolation maps externally defined curves onto the simulation scale. If the raw curve for immunity category \(m\) is specified as values \(\tilde{\gamma}_m(\alpha)\) at high-resolution time indices \(\alpha\), the simulator constructs an interpolated function \(\gamma_m(a)\) compatible with the discrete simulation step index \(a\). The infected agent’s viral load is then:
$v_i(t)=\gamma_{m_i}(t-t_i^{\mathrm{inf}})$.

This representation permits the same mobility and transmission model to be used with different diseases or alternative viral-load profiles.

\subsection{Transmission Mechanism}

Transmission is defined as a local interaction process among co-located agents. At each simulation step, the simulator identifies contact groups induced by shared locations. Within each group, susceptible agents may become infected through exposure to infected agents. This mechanism links the spatial mobility model to the disease-progression model: movement determines which agents meet, while viral load and susceptibility determine whether a contact produces infection.

Let \(C_q(t)\) denote the set of agents located at \(q\in\mathcal{L}\) at step \(t\). A potential transmission event exists for every ordered pair:

\begin{equation}
(i,j)\in C_q(t)\times C_q(t),
\qquad i\neq j,
\end{equation}

\noindent such that agent \(i\) is infected and agent \(j\) is susceptible:

\begin{equation}
z_i(t)=I,\qquad z_j(t)=S.
\end{equation}

The source agent contributes its current viral load \(v_i(t)\), and the target agent contributes its susceptibility \(s_j\). The exposure score for this pair is

\begin{equation}
e_{ij}(t)=v_i(t) \cdot s_j.
\end{equation}

The implemented transmission rule is deterministic. Infection occurs when the exposure score exceeds a fixed threshold \(\theta_{\mathrm{tr}}\):

\begin{equation}
z_j(t^+)=I
\quad \text{if} \quad
e_{ij}(t)>\theta_{\mathrm{tr}}.
\end{equation}

When this condition holds, the susceptible agent becomes infected, its infection time is recorded, and its viral load trajectory begins from the first point of the curve associated with its immunity category. If the condition does not hold, the agent remains susceptible unless another infected agent in the same contact group transmits to it during the same step.

The use of a deterministic threshold rule is a deliberate abstraction. Many epidemic simulations use probabilistic transmission, where exposure determines the probability of infection. A deterministic threshold rule is useful for the present study because the paper focuses on near-critical behavior and counterfactual sensitivity. When all other random choices are fixed by the seed, a small change in mobility can remove a specific contact, prevent a specific infection, and redirect the later trajectory. This makes the causal structure of counterfactual interventions easier to analyze.

As mentioned above, the model supports transmission in both daytime grid cells and homes, but in the experimental configuration used here, transmission is evaluated only for grid cells. 

If multiple infected agents co-locate with the same susceptible agent, the susceptible agent is tested against all relevant infected sources in that location. Under the deterministic threshold rule, any source whose viral load and target susceptibility exceed the threshold is sufficient to infect the target. The model therefore uses a logical $or$ over possible transmitters in the same contact group:

\begin{equation}
z_j(t^+)=I
\quad \text{if } \exists i\in C_q(t)
\text{ such that }
z_i(t)=I
\text{ and }
v_i(t)s_j>\theta_{\mathrm{tr}}.
\end{equation}

The current transmission rule has three useful methodological properties. First, it is local: infection depends only on agents in the same location at the same step. Second, it is heterogeneous: the same transmitter may infect one susceptible agent and not another because susceptibility differs across agents. Third, it is reproducible: once the initial state is fixed, transmission outcomes contain no additional random draw.

The mechanism is intentionally minimal. It does not model cumulative dose over repeated exposures, within-cell distance, contact duration inside a step, environmental persistence, vaccination status, masking, or network ties independent of co-location. These factors could be added by modifying the dose function or the contact rule. The present form was chosen to isolate the dynamical question studied in the paper: whether early aggregate signals can identify near-critical epidemic trajectories and whether small changes in mobility can redirect those trajectories before the system passes into a self-sustaining outbreak regime.

\subsection{Aggregate Observables and Outcome Definitions}

The simulator produces a complete microscopic trajectory, but the prediction and intervention components operate primarily on aggregate epidemic observables. This distinction is important for the methodology. The multi-agent simulator evolves individual states, locations, viral loads, and contacts, while the early warning model receives only compressed population-level summaries. The prediction task is therefore formulated as an inference problem from partial observations rather than from complete knowledge of the microscopic system.

Let \(X_t\) denote the full simulator state at step \(t\). The daily observation vector is obtained through an observation map
$Y_d=\phi(X_{dL:(d+1)L-1})$,
where \(d\) is the day index and \(L\) is the number of simulation steps per day. The map \(\phi\) aggregates step-level events and end-of-day states into a finite-dimensional vector. The main epidemic counts are
$S_d$, $I_d$, $R_d$, $D_d$,
which denote the number of susceptible, infected, recovered, and dead agents at the end of day \(d\). These quantities define the coarse epidemic state visible to the learning model.

Daily event counts are also recorded. Let
$\Delta I_d$, $\Delta R_d$, $\Delta D_d$
denote the number of new infections, new recoveries, and new deaths during day \(d\). These variables capture the local speed of the epidemic process. Two trajectories may have similar values of \(I_d\) at a given time while having different daily incidence patterns. The inclusion of event counts therefore helps distinguish a stable infected population from one that is growing rapidly.

The simulator also records mobility-relevant infected counts, especially the distinction between infected agents who remain mobile and infected agents who are homebound because of disease severity. Let
$I^{\mathrm{mob}}_d$
denote the number of infected agents who are still participating in daytime grid movement at the end of day \(d\), and let
$I^{\mathrm{home}}_d$
denote the number of infected agents restricted to home by the disease process. This distinction is epidemiologically relevant because infectiousness alone does not determine transmission risk. An agent with high viral load can contribute to spread only if the agent participates in contact-generating movement (given the grid-based transmission setting used in the experiments).

The observable vector can therefore be written as:

\begin{equation}
Y_d =
\left(
S_d,
I_d,
R_d,
D_d,
\Delta I_d,
\Delta R_d,
\Delta D_d,
I^{\mathrm{mob}}_d,
I^{\mathrm{home}}_d,
\eta_d
\right),
\end{equation}

\noindent where \(\eta_d\) denotes additional viral-load summaries. These summaries may include quantities such as mean, maximum, or distributional statistics over infected agents’ viral loads. The specification states that the simulator records new infections, current infections, recoveries, deaths, and viral-load summaries at each step, with full statistics available after the simulation.

The final epidemic outcome is defined through the attack rate. For a simulation horizon of \(T\) days, the final attack rate is:
$\rho = {(N-S_T)} / {N}$.
This definition counts the fraction of the population that was no longer susceptible by the end of the run. It therefore includes agents who are infected, recovered, or dead at the final time. In the reported experiments, a trajectory is labeled as a major outbreak if
$\rho \geq \rho_c$,
with
$\rho_c=0.3$.

The corresponding binary outcome label is:

\begin{equation}
y =
\begin{cases}
1, & \rho \geq \rho_c,\\
0, & \rho < \rho_c.
\end{cases}
\end{equation}

Here \(y=1\) denotes a major outbreak and \(y=0\) denotes a contained trajectory.

The early warning problem is defined on a short prefix of the daily observable sequence. Given an observation window of length \(k\):

\begin{equation}
{Y}_{a:b}=(Y_a,Y_{a+1},\ldots,Y_b),
\qquad b-a+1=k,
\end{equation}

\noindent the prediction model estimates the probability that the complete trajectory will become a major outbreak:

\begin{equation}
\hat{p}_{b} =
P(y=1 \mid {Y}_{a:b}).
\end{equation}

The subscript \(b\) emphasizes that the prediction is made at the end of the observed window. In the experiments, this window lies early in the run, before the final epidemic size is known. The learning model must therefore infer the long-term regime from early aggregate signals such as incidence growth, susceptible depletion, active infection counts, and mobility-relevant infected counts.

This formulation supports the central objective of the paper. The model does not merely classify completed outbreaks after the fact. It estimates the risk of a major outbreak while the trajectory is still developing. The intervention procedure can then be applied to high-risk trajectories, especially those near the boundary between contained and outbreak regimes. In such cases, a small counterfactual mobility restriction may alter the contact sequence enough to reduce the final attack rate below the outbreak threshold.

The same outcome definitions are used for baseline and intervention runs. If \(\rho^{0}\) is the final attack rate of the baseline run and \(\rho^{u}\) is the final attack rate under intervention \(u\), the intervention effect on final attack rate is:

\begin{equation}
\Delta \rho(u)=\rho^{0}-\rho^{u}.
\end{equation}

An intervention is considered outbreak-preventing when:

\begin{equation}
\rho^{0}\geq \rho_c
\quad \text{and} \quad
\rho^{u}<\rho_c.
\end{equation}

This definition separates two forms of improvement. A counterfactual intervention may reduce the attack rate without changing the binary outbreak label, or it may move the system across the outbreak threshold. The latter case is the strongest evidence that the original trajectory lay near a tipping boundary.

\subsubsection*{Specific Details}

The simulator records both step-level and daily quantities. The exact set of stored variables may vary across scripts, but the following quantities define the core observable structure used by the prediction and intervention experiments.

\begin{longtable}{@{}p{0.25\textwidth} p{0.1\textwidth} p{0.2\textwidth} p{0.45\textwidth}@{}}
\textbf{Quantity} & \textbf{Symbol} & \textbf{Type} & \textbf{Description} \\
\hline
Susceptible count & \(S_d\) & End-of-day count & Number of agents still susceptible \\
Infected count & \(I_d\) & End-of-day count & Number of currently infected agents \\
Recovered count & \(R_d\) & End-of-day count & Number of recovered agents \\
Death count & \(D_d\) & End-of-day count & Number of dead agents \\
New infections & \(\Delta I_d\) & Daily event count & Number of agents infected during day \(d\) \\
New recoveries & \(\Delta R_d\) & Daily event count & Number of agents recovered during day \(d\) \\
New deaths & \(\Delta D_d\) & Daily event count & Number of agents dying during day \(d\) \\
Mobile infected & \(I^{\mathrm{mob}}_d\) & End-of-day count & Infected agents still participating in daytime movement \\
Homebound infected & \(I^{\mathrm{home}}_d\) & End-of-day count & Infected agents restricted to home by viral load \\
Viral-load summaries & \(\eta_d\) & Summary statistics & Aggregate summaries of infected agents’ \newline viral loads \\
\end{longtable}

The population counts satisfy:

\begin{equation*}
S_d+I_d+R_d+D_d=N
\end{equation*}

\noindent at the end of each day, assuming every agent belongs to exactly one epidemiological state. Daily event counts satisfy:

\begin{equation*}
S_d = S_{d-1}-\Delta I_d
\end{equation*}

\noindent when no other population changes are considered. The recovered and death counts evolve as:

\begin{equation*}
R_d = R_{d-1}+\Delta R_d,
\end{equation*}

\begin{equation*}
D_d = D_{d-1}+\Delta D_d.
\end{equation*}

The infected count follows:

\begin{equation*}
I_d = I_{d-1}+\Delta I_d-\Delta R_d-\Delta D_d.
\end{equation*}

The final attack rate can equivalently be written as:

\begin{equation*}
\rho=\frac{N-S_T}{N}= \frac{I_T+R_T+D_T}{N}.
\end{equation*}

When the epidemic has ended by the final day, \(I_T=0\), the attack rate reduces to:

\begin{equation*}
\rho=\frac{R_T+D_T}{N}.
\end{equation*}

Several derived quantities are useful for analysis:

\begin{longtable}{@{}p{0.38\textwidth} p{0.2\textwidth} p{0.45\textwidth}@{}}
\textbf{Derived outcome} & \textbf{Definition} & \textbf{Interpretation} \\
\hline
Peak infected count & \(\max_d I_d\) & Maximum active infection burden \\
Peak day & \(\arg\max_d I_d\) & Timing of maximum active infection \\
Final susceptible fraction & \(S_T/N\) & Fraction of population never infected \\
Final attack rate & \((N-S_T)/N\) & Fraction ever infected \\
Final mortality fraction & \(D_T/N\) & Fraction dead by final day \\
Cumulative infections & \(\sum_d \Delta I_d\) & Number of secondary and later infections \\
Incidence peak & \(\max_d \Delta I_d\) & Maximum daily new infections \\
Intervention attack rate reduction & \(\rho^0-\rho^u\) & Final-size reduction under intervention \\
Intervention peak reduction & \(\max_d I^0_d-\max_d I^u_d\) & Reduction in maximum active infections \\
\end{longtable}

\subsection{Koopman Representation}

The Koopman operator provides a way to represent nonlinear dynamical systems through the evolution of observables. Consider a deterministic discrete time system:

\begin{equation}
x_{t+1}=F(x_t),
\end{equation}

\noindent where \(x_t\) is the system state and \(F\) is the nonlinear transition map. Instead of acting directly on states, the Koopman operator acts on scalar observables of the state. For an observable:

\begin{equation}
g:\mathcal{X}\rightarrow \mathbb{R},
\end{equation}

\noindent the Koopman operator \(\mathcal{K}\) is defined as:

\begin{equation}
(\mathcal{K}g)(x)=g(F(x)).
\end{equation}

Although the state transition \(F\) may be nonlinear, the Koopman operator is linear on the space of observables:

\begin{equation}
\mathcal{K}(a g_1+b g_2)=a\mathcal{K}g_1+b\mathcal{K}g_2.
\end{equation}

This property is useful because many complex systems are difficult to linearize in their original state variables. A suitable set of observables may evolve approximately according to a simpler linear rule, even when the underlying system remains nonlinear.

In the epidemic simulation, the full microscopic state \(X_t\) includes all agent locations, disease states, viral loads, and internal timers. The early warning model does not receive this full state. It receives daily aggregate observables $Y_d$. The Koopman approximation is therefore learned over the observed epidemic trajectory rather than over the complete simulator state. The goal is to construct a lifted representation:

\begin{equation}
z_d=\psi({Y}_{d-k+1:d})
\end{equation}

\noindent where \(\psi\) maps an observed history window into a latent space. In that latent space, the dynamics are approximated by a linear transition matrix \(A\):

\begin{equation}
z_{d+1}\approx A z_d.
\end{equation}

Multi-step prediction then follows from repeated application of the same linear operator:

\begin{equation}
z_{d+\ell}\approx A^\ell z_d,
\qquad \ell=1,\ldots,h.
\end{equation}

A decoder or reconstruction map \(\Gamma\) maps latent states back to observable space:

\begin{equation}
\widehat{Y}_{d+\ell}=\Gamma(A^\ell z_d).
\end{equation}

This formulation contains the main idea behind practical Koopman learning. The lifted space is not given in advance, but it must be computed from data. Classical methods often choose a dictionary of basis functions, such as polynomials, radial basis functions, or other hand-designed nonlinear transformations. Modern AI implementations more often learn the lifting map \(\psi\) directly, typically with a neural encoder or another flexible representation model. The learned representation is trained so that the encoded trajectory can be advanced approximately by a linear operator and decoded back into the observed variables.

A typical training objective contains three related terms. The first encourages the latent representation to reconstruct the observed epidemic variables:

\begin{equation}
\mathcal{L}_{\mathrm{rec}}=
\sum_d
\left\|
\Gamma(\psi({Y}_{d-k+1:d}))-Y_d
\right\|^2.
\end{equation}

The second encourages one-step linear evolution in latent space:

\begin{equation}
\mathcal{L}_{\mathrm{lin}}=
\sum_d
\left\|
\psi({Y}_{d-k+2:d+1})
-
A\psi({Y}_{d-k+1:d})
\right\|^2.
\end{equation}

The third encourages accurate multi-step prediction after repeated application of the linear operator:

\begin{equation}
\mathcal{L}_{\mathrm{pred}}=
\sum_d\sum_{\ell=1}^{h}
\left\|
Y_{d+\ell}
-
\Gamma(A^\ell \psi({Y}_{d-k+1:d}))
\right\|^2.
\end{equation}

The overall training objective can be written as:

\begin{equation}
\mathcal{L}=
\mathcal{L}_{\mathrm{rec}}
+
\lambda_{\mathrm{lin}}\mathcal{L}_{\mathrm{lin}}
+
\lambda_{\mathrm{pred}}\mathcal{L}_{\mathrm{pred}},
\end{equation}

\noindent where the \(\lambda\) terms control the relative importance of latent linearity and forecast accuracy.

This approximation does not require the learned coordinates to correspond to directly interpretable epidemiological compartments. Their role is to provide a compact state in which the early trajectory has predictive capability. In the present study, this is especially relevant near the transition between contained and outbreak regimes. A short observed history may contain weak but informative signals about whether infection growth is accelerating, whether susceptible decrease has begun, and whether mobile infected agents remain capable of generating new contacts. The Koopman latent state is trained to preserve such dynamical information because it must support both reconstruction and forward prediction.

The Koopman model used in the implementation follows this general data-driven approximation strategy. It learns a low-dimensional latent representation from simulation trajectories, estimates a linear evolution rule in that latent space, and uses the learned dynamics to forecast aggregate epidemic observables over a short future horizon.

The appeal of the Koopman representation in this setting lies in its balance between structure and flexibility. Purely mechanistic prediction would require detailed use of the full multi-agent state, which is high-dimensional and difficult to summarize. A generic black-box classifier can predict outbreak labels, but it may provide limited information about trajectory evolution. The Koopman component imposes a dynamical constraint through latent linear evolution, while still learning the lifted representation from data. This makes it suitable for early warning analysis in a complex simulation where the original dynamics are nonlinear and individual-based, but the prediction task is defined over aggregate observables.

\subsection{Early Warning Model}

The early warning layer transforms short initial epidemic trajectories into features used to classify whether the full simulation will cross the major outbreak threshold. The simulator itself evolves a high-dimensional microscopic state, but the prediction model operates on daily aggregate observables. This design reflects the intended role of the predictive component: it should infer the long-term epidemic regime from early population-level signals, before the final attack rate is known and before interventions become ineffective.

Let
$Y_d \in \mathbb{R}^{m}$
denote the aggregate observable vector at day \(d\), as defined above. For an early warning decision made at day \(b\), the model receives a fixed-length history:

\begin{equation}
{Y}_{b-k+1:b}=(Y_{b-k+1},Y_{b-k+2},\ldots,Y_b),
\end{equation}

\noindent where \(k\) is the observation-window length. 

In the reported experiments, $k=5$.
Thus, each prediction is based on five consecutive days of aggregate epidemic information. The window may contain population counts, daily incidence variables, mobility-relevant infected counts, and viral load summaries. These variables describe both the current epidemic burden and its short-term temporal change.

The target variable is the final outbreak label
$y=\mathbf{1}[\rho\geq \rho_c]$,
where \(\rho\) is the final attack rate and \(\rho_c=0.3\) is the major-outbreak threshold. The early warning problem is therefore:

\begin{equation}
\hat{p}_b = P(y=1\mid {Y}_{b-k+1:b}),
\end{equation}

\noindent where \(\hat{p}_b\) is the estimated probability that the simulation will end as a major outbreak. The central difficulty is that the observation window is short relative to the full simulation horizon. At early times, contained and outbreak trajectories may still have similar numbers of infected agents, while their future evolution may differ sharply because of contact timing, susceptibility, and viral load progression.

The predictive layer combines two modeling components. The first is a Koopman-inspired dynamical model that learns a low-dimensional latent representation of early epidemic trajectories. The second is a random forest (RF) classifier that maps early window features, optionally including Koopman-derived features, to an outbreak probability. The two components serve different roles. The Koopman model provides a representation of epidemic dynamics and short-horizon forecasts. The random forest provides a robust supervised classifier over heterogeneous features extracted from the early trajectory.

In the reported configuration, the Koopman model uses a five-day observation window and a five-day prediction horizon. For a decision day \(b\), the model encodes the observed window into a latent state
$z_b \in \mathbb{R}^{r}$,
where \(r\) is the latent dimension. The experimental setting uses a low-dimensional latent state, with \(r=6\). This latent state is then evolved forward for \(h=5\) days by the learned Koopman dynamics. The resulting latent trajectory supports short-horizon forecasts of aggregate epidemic variables and provides features for downstream outbreak classification.

The random forest classifier receives a feature vector:

\begin{equation}
x_b=\psi({Y}_{b-k+1:b}, z_b, \widehat{{Y}}_{b+1:b+h}),
\end{equation}

\noindent where \(\psi\) denotes the feature-construction map, \(z_b\) is the Koopman latent representation, and \(\widehat{{Y}}_{b+1:b+h}\) denotes the Koopman forecast over the prediction horizon. Depending on the experiment, the feature vector may include only direct statistical features from the early window, or it may also include Koopman-derived features such as latent coordinates, predicted future counts, or a learned boundary score. The classifier output is:

\begin{equation}
\hat{p}_b=f_{\mathrm{RF}}(x_b),
\end{equation}

\noindent where \(f_{\mathrm{RF}}\) is the trained random forest. A high value of \(\hat{p}_b\) indicates that the observed early trajectory is likely to end in a major outbreak.

The early warning model is designed for use before the counterfactual intervention search. Its function is not only to label completed simulations; it identifies trajectories whose early dynamics suggest an increased probability of crossing the final outbreak threshold. These trajectories can then be examined through intervention experiments. For the study of tipping points, prediction identifies trajectories near dangerous regimes, while counterfactual simulation tests whether small perturbations can move them into a contained regime.

\subsubsection*{Specific Details}

The early warning experiments use the following configuration:

\begin{longtable}{@{}p{0.25\textwidth} p{0.1\textwidth} p{0.1\textwidth} p{0.5\textwidth}@{}}
\textbf{Component} & \textbf{Symbol} & \textbf{Value} & \textbf{Role} \\
\hline
Observation window & \(k\) & 5 days & Number of observed days used for prediction \\
Koopman forecast \newline horizon & \(h\) & 5 days & Number of future days predicted by Koopman model \\
Latent dimension & \(r\) & 6 & Dimension of Koopman latent state \\
Outcome threshold & \(\rho_c\) & 0.3 & Final attack rate threshold for major outbreak \\
Classifier & \(f_{\mathrm{RF}}\) & Random forest & Maps early window features to outbreak \newline probability \\
\end{longtable}

The random forest is used because the final classification boundary may depend on nonlinear combinations of early counts, short-term changes, and Koopman-derived features. It also handles mixed feature types and does not require the classifier itself to impose a parametric epidemic form. The Koopman model and the random forest therefore play complementary roles: the Koopman component supplies a dynamical representation and short-term forecast, while the random forest converts early warning information into an outbreak risk score.

\subsection{Intervention Model}

The intervention component involves counterfactual simulations. A baseline epidemic trajectory is first generated with fixed model parameters, initialization, and random seed. An intervention trajectory is then generated from the same initial conditions, with a single change to agent mobility. The difference between the two trajectories measures the effect of the intervention for the simulated system.

Let
$X^{0}_{0:T}$
denote the baseline trajectory over a horizon of \(T\) days, and let
$X^{u}_{0:T}$
represent the trajectory produced under intervention \(u\). Both trajectories use the same parameter vector \(\theta\), the same initialized population, the same movement routines, the same susceptibility and immunity assignments, and the same initially infected agent. The intervention changes only the movement rule for one selected agent on one selected day. 

The intervention is defined by an agent-day pair
$u=(a,d^\star)$,
where \(a\in\mathcal{A}\) is the selected agent and \(d^\star\in\{0,\ldots,T-1\}\) is the intervention day. Under this intervention, agent \(a\) is quarantined, i.e., forced to remain at home throughout day \(d^\star\). This is a minimal intervention, a very small perturbation of the system. For all other agents and all other days, the simulation follows the baseline rules. In terms of the intervention indicator introduced earlier:

\begin{equation}
q_i(t;u)=
\begin{cases}
1, & i=a \ \text{and} \ d(t)=d^\star,\\
0, & \text{otherwise}
\end{cases}
\end{equation}

\noindent where
$d(t)=\left\lfloor {t}/{L}\right\rfloor$.

The location rule for the selected agent becomes:

\begin{equation}
\ell_a(t)=h_a
\quad
\text{for all } t \text{ such that } d(t)=d^\star.
\end{equation}

After that day, the agent returns to the ordinary mobility pattern. The intervention does not change viral load progression, susceptibility, immunity, infection thresholds, recovery rules, or the movement routines of other agents.

The effect of a one-day movement restriction can propagate through later infections. Preventing one infection can remove an entire downstream infection chain; alternatively, preventing one contact may have little effect if the same susceptible agent is infected later by another source. The intervention effect is therefore a property of the full nonlinear trajectory, not only of the immediate contacts removed on the intervention day.

Several outcome measures are used to quantify intervention effects. Let \(\rho^0\) denote the final attack rate of the baseline trajectory, and let \(\rho^u\) denote the final attack rate under intervention \(u\). 

The attack rate reduction is:

\begin{equation}
\Delta \rho(u)=\rho^0-\rho^u.
\end{equation}

A positive value indicates that the intervention reduced the final fraction of agents infected during the simulation. 

The peak infected count is:

\begin{equation}
P^0=\max_{0\leq d\leq T} I^0_d
\end{equation}

\noindent for the baseline trajectory and:

\begin{equation}
P^u=\max_{0\leq d\leq T} I^u_d
\end{equation}

\noindent for the intervention trajectory. 

The peak reduction is:

\begin{equation}
\Delta P(u)=P^0-P^u.
\end{equation}

This quantity measures whether the intervention reduced the maximum active infection burden, even when the final attack rate remains above the outbreak threshold.

The strongest intervention outcome is outbreak prevention. For a major outbreak threshold \(\rho_c\), an intervention \(u\) is considered outbreak-preventing when
$\rho^0\geq \rho_c$ and $\rho^u<\rho_c$.

This definition captures cases in which the baseline trajectory crosses the outbreak threshold while the counterfactual trajectory remains below it. Such cases are especially relevant for tipping point analysis because they indicate that the baseline trajectory was close enough to the regime boundary for a small perturbation to alter the qualitative outcome.

The search over interventions evaluates a finite set of candidate agent-day pairs,
$\mathcal{U}\subseteq \mathcal{A}\times\{0,\ldots,T-1\}$.
For each candidate \(u\in\mathcal{U}\), the simulator generates a counterfactual trajectory and computes the corresponding outcomes. The best intervention can be selected according to final attack rate reduction:

\begin{equation}
u^\star_{\rho}=
\arg\max_{u\in\mathcal{U}}
\Delta \rho(u),
\end{equation}

\noindent or according to peak infected reduction:

\begin{equation}
u^\star_{P}=
\arg\max_{u\in\mathcal{U}}
\Delta P(u).
\end{equation}

In the reported experiments, final attack rate reduction is the primary criterion because the major outbreak label is defined through final attack rate. Peak reduction is still informative because it captures changes in epidemic burden and timing.

The role of the intervention is to test whether the simulated epidemic trajectory is modifiable near the outbreak boundary. When a single agent-day perturbation prevents a major outbreak in the model, the result indicates the presence of a sensitive dynamical region in the simulated system. This counterfactual procedure connects the intervention model to the early warning layer. The early warning model identifies trajectories likely to enter the outbreak regime from short observation windows. The intervention search then tests whether such trajectories can still be redirected. In near-critical regimes, a small change in the contact sequence may prevent the infection of a high-impact agent or interrupt an early transmission chain. In regimes far from the boundary, the same intervention may have negligible effect because the trajectory is already strongly directed toward extinction or widespread outbreak.

\subsubsection*{Specific Details}

For each intervention candidate, the following metrics are computed:

\begin{longtable}{@{}p{0.3\textwidth} p{0.2\textwidth} p{0.5\textwidth}@{}}
\textbf{Metric} & \textbf{Definition} & \textbf{Interpretation} \\
\hline
Baseline attack rate & \(\rho^0\) & Final fraction infected without intervention \\
Intervention attack rate & \(\rho^u\) & Final fraction infected under intervention \\
Attack rate reduction & \(\Delta\rho(u)=\rho^0-\rho^u\) & Reduction in final epidemic size \\
Baseline peak infected & \(P^0=\max_d I^0_d\) & Maximum active infections without intervention \\
Intervention peak infected & \(P^u=\max_d I^u_d\) & Maximum active infections under intervention \\
Peak reduction & \(\Delta P(u)=P^0-P^u\) & Reduction in maximum active infections \\
Outbreak prevention & \(\rho^0\geq\rho_c,\ \rho^u<\rho_c\) & Whether the intervention changes the binary outcome \\
\end{longtable}

\section{Experimental Results}
\setcounter{table}{0}

\begin{figure}[!htbp]
\centering
\begin{tikzpicture}[
    >=Latex,
    block/.style={
        draw,
        rounded corners,
        align=center,
        text width=3.0cm,
        minimum height=1.0cm,
        inner sep=4pt,
        font=\footnotesize
    },
    arrow/.style={->, thick}
]

% Top row
\node[block] (sim) at (0,0) {Epidemic\\simulation};
\node[block] (obs) at (4.6,0) {Early aggregate\\observations};
\node[block] (koop) at (9.2,0) {Koopman latent\\representation};

% Bottom row
\node[block] (out) at (0,-2.2) {Outcome comparison\\and selected minimal intervention};
\node[block] (int) at (4.6,-2.2) {Counterfactual\\intervention test};
\node[block] (warn) at (9.2,-2.2) {Early outbreak\\warning};

% Arrows
\draw[arrow] (sim) -- (obs);
\draw[arrow] (obs) -- (koop);
\draw[arrow] (koop) -- (warn);
\draw[arrow] (warn) -- (int);
\draw[arrow] (int) -- (out);

\end{tikzpicture}
\caption{Overview of the proposed workflow. Epidemic simulations generate early aggregate observations, which are mapped into a Koopman latent representation for early outbreak warning. High-risk trajectories are then evaluated with counterfactual interventions to identify small changes that may reduce the outbreak.}
\label{fig2-koopman-epidemic-workflow}
\end{figure}
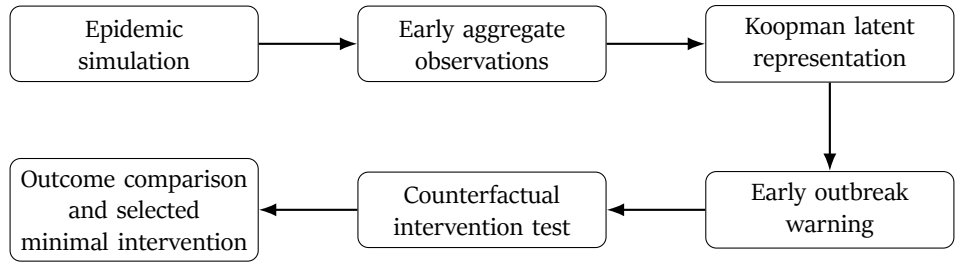

The experiments presented in this section evaluate whether early aggregate trajectories from the multi-agent epidemic simulator contain enough information to predict the final epidemic regime and whether trajectories predicted to be dangerous can still be changed by a minimal counterfactual intervention. The evaluation has several parts. The simulation ensemble is constructed near the transition between contained and major outbreak behavior. A Koopman model is trained to encode short aggregate histories into a low-dimensional latent state with approximately linear evolution. An early warning classifier maps direct early window statistics and Koopman-derived quantities to a probability of major outbreak. Finally, selected outbreak candidates are rerun under single-agent, single-day quarantine interventions.
The entire system workflow is displayed in Figure~\ref{fig2-koopman-epidemic-workflow}.

\subsection{Boundary-Focused Simulation Dataset}

The dataset was generated to focus near the transition between rapid extinction and sustained outbreak. Empirically, the lower susceptibility bound was fixed at $1.3$, while the upper bound was swept from $1.3015$ to $1.3025$ in increments of $2\times 10^{-5}$. For each upper bound value, 200 random seeds were simulated, giving 10,200 runs. The aggregate trajectory stored for each run contains nine daily observables: susceptible agents, infected agents, recovered agents, dead agents, new infections, new recoveries, new deaths, infected mobile agents, and infected homebound agents. These observables are the input to both the Koopman model and the supervised early warning classifier. Some details are presented in Table~\ref{tab:dataset-summary}.

\begin{table}[htbp]
\centering
\caption{Details about the simulation dataset and early window construction.}
\label{tab:dataset-summary}
\begin{tabular}{ll}
\toprule
\textbf{Quantity} & \textbf{Value} \\
\midrule
Simulation runs & 10,200 \\
Agents per run & 500 \\
Random seeds & 200 seeds \\
Susceptibility lower bound & 1.3 \\
Susceptibility upper-bound values & 51 values, 1.3015--1.3025 \\
Dataset maximum horizon & 365 days with early stopping \\
Major outbreak threshold & Final attack rate $\rho \geq 0.3$ \\
Contained runs & 4,825 \\
Major outbreak runs & 5,375 \\
Stored observables & 9 daily aggregate variables \\
Koopman history window & 5 days \\
Koopman forecast horizon & 5 days \\
Early window end days & 4--12 \\
Koopman latent dimension & 6 \\
Train/validation/test runs & 7,140 / 1,530 / 1,530 \\
Train/validation/test Koopman windows & 49,719 / 10,548 / 10,772 \\
\bottomrule
\end{tabular}
\end{table}

Table~\ref{tab:regime-summary} summarizes the two regimes obtained from the final attack rate label. The dataset is nearly balanced, but the two outcome classes are sharply separated in final size. Contained runs infect between $0.2\%$ and $2.4\%$ of the population, whereas major outbreaks infect between $59\%$ and $83\%$. This separation supports the binary major outbreak label. At the same time, the narrow susceptibility sweep keeps the system close to a transition region, since similar parameter values and different random seeds can produce qualitatively different outcomes.

\begin{table}[htbp]
\centering
\caption{Empirical regime statistics in the boundary-focused dataset. Attack rates are fractions of the population.}
\label{tab:regime-summary}
\setlength{\tabcolsep}{4pt}
\begin{tabular}{@{}p{0.18\textwidth} p{0.09\textwidth} p{0.14\textwidth} p{0.13\textwidth} p{0.15\textwidth} p{0.13\textwidth}@{}}
\toprule
\textbf{Regime} & \textbf{Runs} & \textbf{Attack rate range} & \textbf{Mean \newline attack \newline rate} & \textbf{Peak \newline infected \newline range} & \textbf{Mean \newline peak \newline infected} \\
\midrule
Contained & 4,825 & 0.002--0.024 & 0.004 & 1--12 & 2.0 \\
Major outbreak & 5,375 & 0.592--0.830 & 0.726 & 78--232 & 156.7 \\
\bottomrule
\end{tabular}
\end{table}

\subsection{Koopman Representation of Early Trajectories}

\begin{figure}[htbp]
\centering
\includegraphics[width=0.95\linewidth]{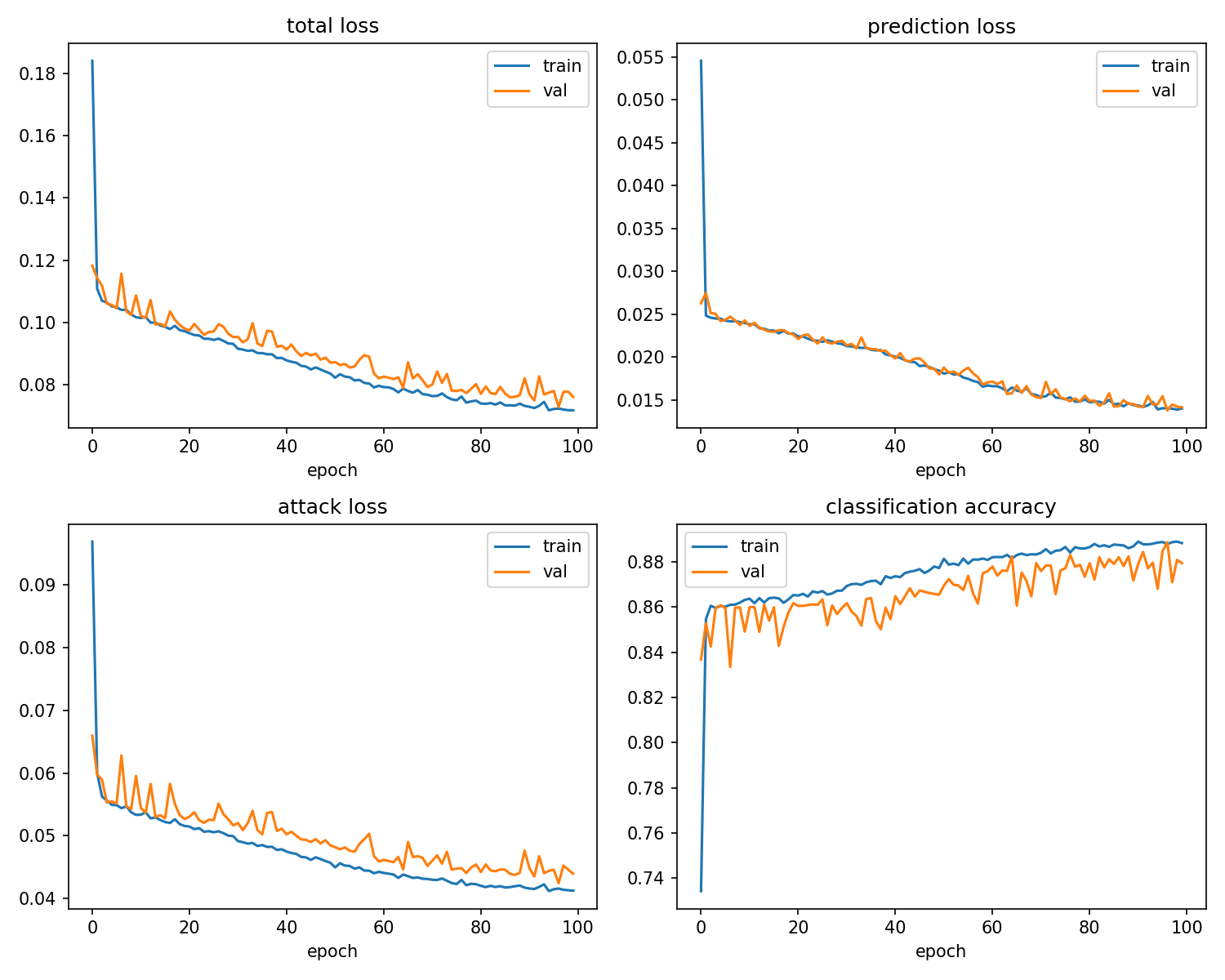}
\caption{Koopman training curves for total loss, prediction loss, attack rate loss, and classification accuracy.}
\label{fig:koopman-training-curves}
\end{figure}

The Koopman model was trained on five day observation windows and optimized to predict the next five days of aggregate dynamics while maintaining approximately linear latent evolution. The model uses a six-dimensional latent state. Model selection was based on validation loss. Figure~\ref{fig:koopman-training-curves} displays the model training curves, while Table~\ref{tab:koopman-training-summary} shows quantitative details about the training process.

\begin{table}[htbp]
\centering
\caption{Koopman model configuration and selected training losses.}
\label{tab:koopman-training-summary}
\begin{tabular}{lrr}
\toprule
\textbf{Quantity} & \textbf{Training} & \textbf{Validation} \\
\midrule
Windows & 49,719 & 10,548 \\
Runs & 7,140 & 1,530 \\
Total loss & 0.0723 & 0.0729 \\
Forecast loss & 0.0140 & 0.0138 \\
Latent linearity loss & 0.0038 & 0.0035 \\
Attack rate loss & 0.0416 & 0.0424 \\
Classification loss & 0.2453 & 0.2486 \\
Classification accuracy & 0.8879 & 0.8887 \\
\bottomrule
\end{tabular}
\end{table}

\begin{figure}[!htbp]
\centering
\includegraphics[width=0.88\linewidth]{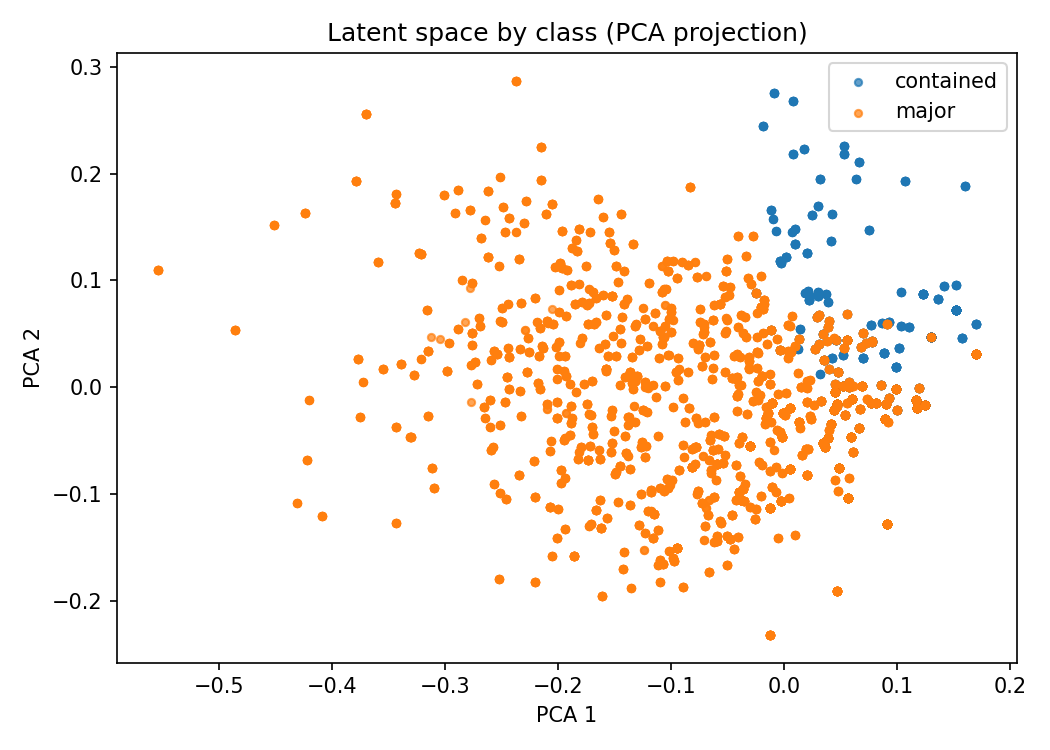}
\caption{PCA projection of six-dimensional Koopman latent states in the test set, colored by final outbreak class.}
\label{fig:latent-scatter-class}
\end{figure}

\begin{figure}[!htbp]
\centering
\includegraphics[width=0.88\linewidth]{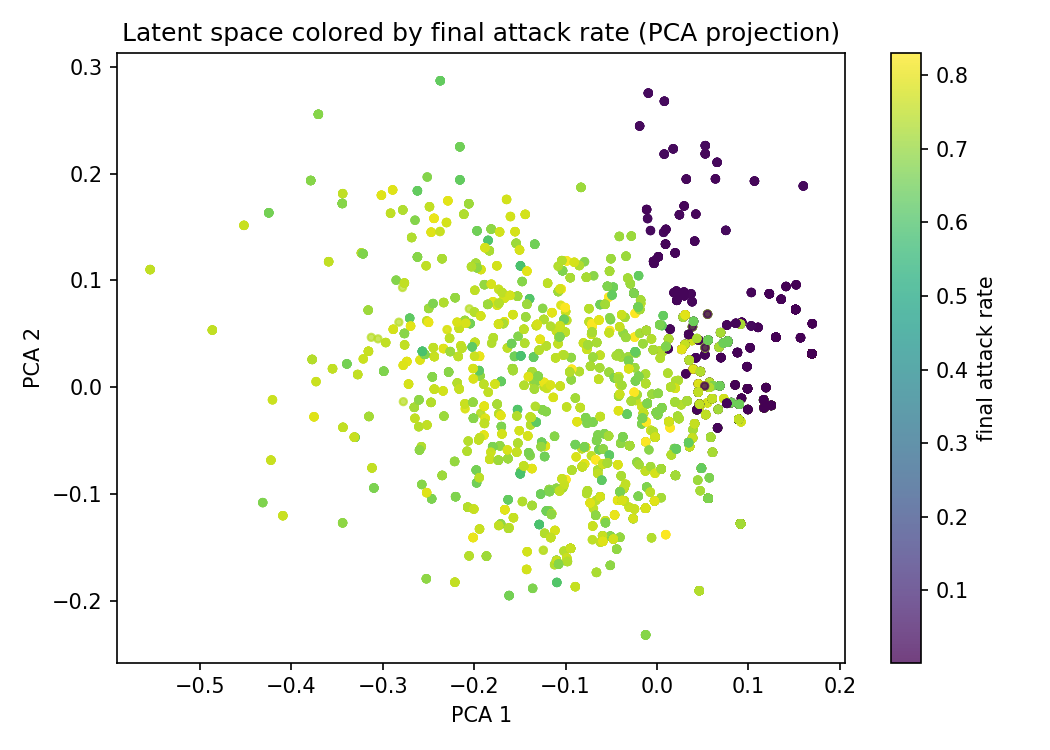}
\caption{PCA projection of the same test set Koopman latent states, colored by final attack rate.}
\label{fig:latent-scatter-attack-rate}
\end{figure}

The learned latent state contains outcome-related information. In the PCA projection of the test latent states presented in Figure~\ref{fig:latent-scatter-class} and Figure~\ref{fig:latent-scatter-attack-rate}, contained and major outbreak windows occupy different regions of the latent space, and coloring by final attack rate shows a continuous organization by eventual epidemic severity. These plots provide visual evidence that the six-dimensional Koopman representation preserves information relevant to long-term outcomes.

\begin{figure}[!htbp]
\centering
\includegraphics[width=0.95\linewidth]{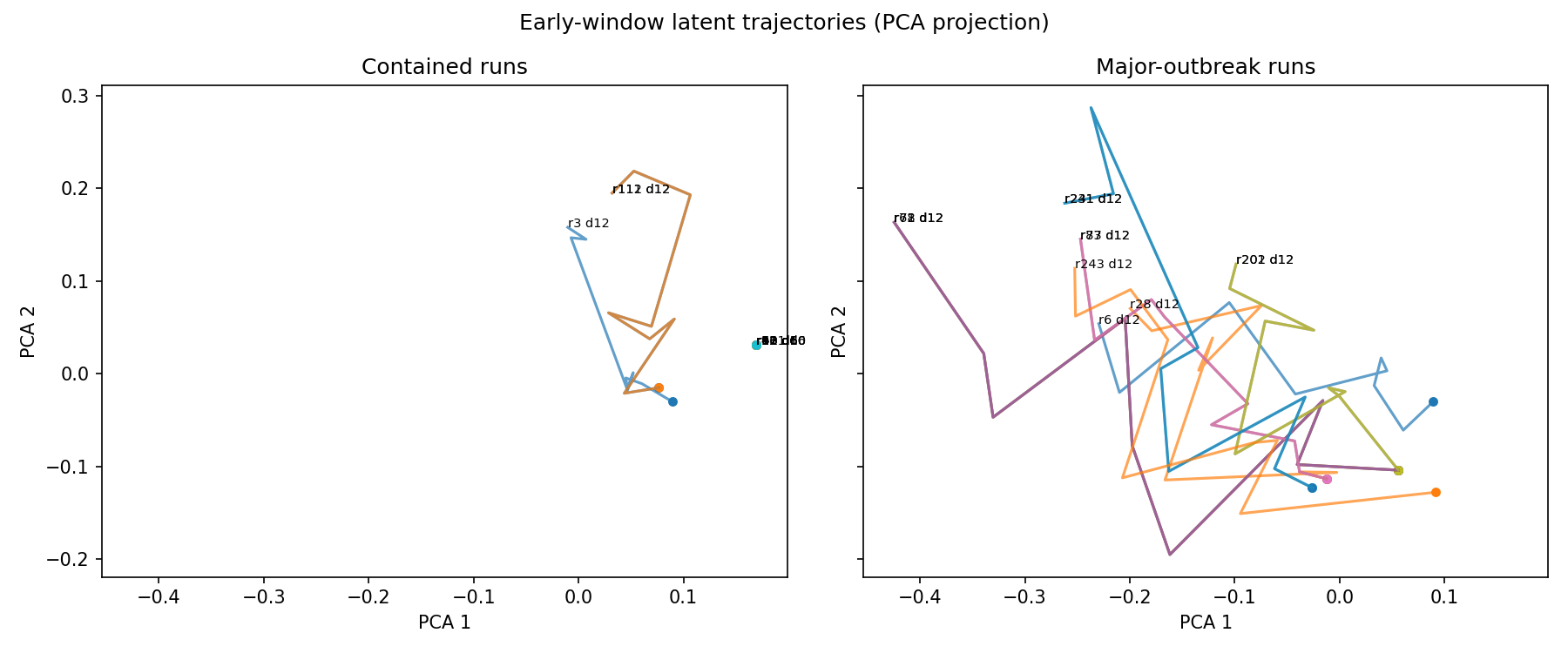}
\caption{Representative early window trajectories in the Koopman latent space, shown after PCA projection. The trajectories show different regions for contained and major outbreak runs.}
\label{fig:latent-trajectories}
\end{figure}

Figure~\ref{fig:latent-trajectories} shows representative early window trajectories in the learned Koopman latent space after projection onto the first two principal components. Each curve corresponds to the evolution of one simulation window, with labels marking the run and day index. The contained trajectories occupy a compact region of the projected space and show limited dispersion, which suggests that their early dynamics remain close to a stable, low-growth regime. In contrast, the major outbreak trajectories cover a broader portion of the latent plane and exhibit more variable directions of motion. This larger spread indicates that outbreak trajectories are not simply distinguished by a single scalar increase in infection count, but by a richer dynamical pattern captured by the latent representation. The separation between the two cases supports the use of Koopman features as early indicators of the final epidemic regime.

The Koopman-derived probability for a major outbreak also provides a direct early warning signal. As shown in Table~\ref{tab:koopman-test-performance}, at the window level, thresholding this probability at 0.5 gives 0.8886 accuracy and 0.9558 ROC AUC (Area Under the Receiver Operating Characteristic Curve) on 10,772 test windows. At the run level, using the last available early window probability gives stronger discrimination: accuracy rises to 0.9732 and ROC AUC to 0.9980. This difference is expected because later early windows contain more information than the first eligible windows, while the run-level summary suppresses within-run noise.

\begin{table}[!htbp]
\centering
\caption{Test set performance of the Koopman-derived major outbreak probability at threshold 0.5.}
\label{tab:koopman-test-performance}
\begin{tabular}{lrr}
\toprule
\textbf{Metric} & \textbf{Window level} & \textbf{Run level} \\
\midrule
Test units & 10,772 & 1,530 \\
Contained units & 3,428 & 714 \\
Major outbreak units & 7,344 & 816 \\
Accuracy & 0.8886 & 0.9732 \\
ROC AUC & 0.9558 & 0.9980 \\
Average precision & 0.9789 & 0.9982 \\
True contained predicted contained & 2,870 & 706 \\
True contained predicted major & 558 & 8 \\
True major predicted contained & 642 & 33 \\
True major predicted major & 6,702 & 783 \\
\bottomrule
\end{tabular}
\end{table}

\subsection{Early Warning Random Forest Classifier}

The supervised early warning classifier estimates whether a trajectory will end as a major outbreak. Each prediction is made from a five-day observation window of aggregate epidemic variables. The input features combine direct summaries of the observed time series with Koopman-derived quantities from the learned latent representation. The train-test split is performed by simulation run, so windows from the same trajectory are not shared across partitions.

Table~\ref{tab:rf-summary} summarizes the held-out performance. The classifier achieves very high discrimination on the test windows, with accuracy above $0.99$ and near-perfect ranking metrics. Precision and recall are also balanced across the contained and major outbreak classes, which indicates that the classifier is not obtaining its performance by favoring one class. These results show that the early aggregate trajectories contain strong information about the final epidemic regime in the boundary-focused simulation.

\begin{table}[!htbp]
\centering
\caption{Random forest early warning classifier performance on held-out test windows.}
\label{tab:rf-summary}
\begin{tabular}{lr}
\toprule
\textbf{Quantity} & \textbf{Value} \\
\midrule
Observation window & 5 days \\
Early window end days & 4--12 \\
Held-out test windows & 17,654 \\
Test accuracy & 0.9958 \\
ROC AUC & 0.9999 \\
Average precision & 0.9999 \\
Contained precision / recall & 0.9933 / 0.9977 \\
Major outbreak precision / recall & 0.9979 / 0.9940 \\
\bottomrule
\end{tabular}
\end{table}

The temporal breakdown in Table~\ref{tab:rf-by-end-day} shows how predictive information accumulates during the early phase of the epidemic. Even the earliest available windows, ending on day~4, already provide strong discrimination. Performance improves as additional days are observed, and the stored test predictions become perfectly separated from day~8 onward. This pattern is consistent with the intended early warning role of the classifier: the first few days contain useful but incomplete information, while subsequent early observations make the final regime increasingly explicit.

\begin{table}[!htbp]
\centering
\caption{Random forest test performance by end day of the five-day observation window.}
\label{tab:rf-by-end-day}
\begin{tabular}{rrrr}
\toprule
\textbf{End day} & \textbf{Test windows} & \textbf{Accuracy} & \textbf{ROC AUC} \\
\midrule
4 & 2,040 & 0.9755 & 0.9974 \\
5 & 2,040 & 0.9926 & 0.9999 \\
6 & 2,040 & 0.9975 & 1.0000 \\
7 & 2,040 & 0.9975 & 1.0000 \\
8 & 2,040 & 1.0000 & 1.0000 \\
9 & 2,040 & 1.0000 & 1.0000 \\
10 & 2,040 & 1.0000 & 1.0000 \\
11 & 2,040 & 1.0000 & 1.0000 \\
12 & 1,334 & 1.0000 & 1.0000 \\
\bottomrule
\end{tabular}
\end{table}

Table~\ref{tab:rf-feature-families} groups the most relevant features by source. The largest contributions come from infected-series features and Koopman-derived features. This is expected: the active infected population describes the current epidemic burden, while the Koopman variables summarize short-horizon dynamical evolution in the learned latent space. Incidence series and susceptible series features also contribute substantially, reflecting the importance of early growth and susceptible depletion. The much smaller contribution of susceptibility parameters indicates that the classifier relies primarily on trajectory-derived information.

\begin{table}[!htbp]
\centering
\caption{Random forest feature importance grouped by feature family.}
\label{tab:rf-feature-families}
\begin{tabular}{lr}
\toprule
\textbf{Feature family} & \textbf{Importance} \\
\midrule
Infected series features & 0.2357 \\
Koopman-derived features & 0.2249 \\
Incidence series features & 0.1826 \\
Susceptible series features & 0.1491 \\
Susceptibility parameters & 0.0080 \\
\bottomrule
\end{tabular}
\end{table}

The random forest results support the use of early aggregate trajectories for outbreak warning in the simulated near-boundary regime. The feature family analysis further shows that Koopman-derived quantities provide information comparable in magnitude to the strongest direct epidemic summaries, which supports their role as compact dynamical descriptors rather than only auxiliary forecast outputs.

\subsection{Counterfactual Quarantine Intervention}

The counterfactual experiments test whether trajectories that develop into major outbreaks can still be redirected by a minimal mobility restriction. The random forest classifier is used only as a prioritization step: trajectories with high outbreak risk are selected for full baseline verification, and only trajectories that actually cross the outbreak threshold in the complete baseline simulation are passed to the intervention search. The final intervention results are therefore evaluated against the baseline outcome, not against the classifier prediction.

Each candidate intervention is an agent-day pair $u=(a,d)$. Under this intervention, the selected agent is quarantined at home for the entire day $d$, and the simulation is rerun from the same initialized population, mobility routines, susceptibility values, immunity categories, and initially infected agent. The only difference between the baseline and intervention runs is the one-day mobility restriction. As mentioned before, the intervention effect is measured by the final attack rate reduction
$ \Delta \rho(u) = \rho_0 - \rho_u$,
where $\rho_0$ is the baseline final attack rate and $\rho_u$ is the final attack rate under intervention. An intervention is counted as outbreak-preventing when $\rho_0 \geq 0.3$ and $\rho_u < 0.3$.

The six cases reported in this subsection are illustrative counterfactual scenarios. They are not intended to estimate the overall success rate of the intervention procedure. The simulator can generate any number of baseline trajectories and evaluate any number of candidate interventions, and the frequency of successful prevention depends on the sampled parameter regime, the intervention search space, and the outbreak threshold. The purpose of these examples is to show the main qualitative outcomes that arise: complete prevention, partial mitigation, and delay without prevention.

Table~\ref{tab:intervention-summary} summarizes these qualitative outcomes. In the prevented examples, the selected one-day quarantine moves the trajectory from a major outbreak to near extinction: the average final attack rate falls from approximately $0.69$ to almost zero, and the active infection peak is reduced by more than 100 agents. The two non-prevented examples show that a minimal intervention is not always sufficient to cancel the outbreak. In one case, the intervention mainly delays the epidemic wave. In the other, it reduces the maximum active burden but leaves the final attack rate above the major outbreak threshold. These outcomes are important because they show that the same intervention type can have different effects depending on the dynamical position of the trajectory.

\begin{table}[!htbp]
\centering
\small
\caption{Aggregate summary of the illustrative counterfactual one-day quarantine outcomes. Peak reduction is defined as $\Delta P=P^0-P^u$, where $P^0$ and $P^u$ are the baseline and intervention peak infected counts. The case counts describe only the examples reported in this subsection and should not be interpreted as an estimate of overall intervention success probability.}
\label{tab:intervention-summary}
\begin{tabular}{lrrrrr}
\toprule
\textbf{Outcome type} &  $\boldsymbol{\rho^0}$ & $\boldsymbol{\rho^u}$ & $\boldsymbol{\Delta\rho}$ & $\boldsymbol{\Delta P}$ \\
\midrule
Prevented outbreak  & 0.691 & 0.004 & 0.688 & 130.5 \\
Delayed outbreak  & 0.710 & 0.690 & 0.020 & 0 \\
Reduced peak &  0.784 & 0.756 & 0.028 & 51.0 \\
\bottomrule
\end{tabular}
\end{table}

Table~\ref{tab:intervention-case-summary} presents the same examples at the case level. The table gives the intervention day, the baseline and intervention peak infected counts, the timing of those peaks, the final attack rates, and the qualitative outcome. Cases 1, 3, 4, and 6 show complete outbreak prevention: the baseline trajectories cross the major outbreak threshold, while the intervention trajectories remain far below it. Case~2 shows a delayed outbreak: the peak occurs much later after intervention, but the final attack rate remains high. Case~5 shows partial mitigation: the intervention reduces and delays the peak, but the final epidemic size remains in the major outbreak regime.

\begin{table}[!htbp]
\centering
\caption{Summary of the counterfactual one-day quarantine interventions. Attack rates are final fractions of the population infected over the simulation horizon.}
\label{tab:intervention-case-summary}
\small
\setlength{\tabcolsep}{4pt}
\begin{tabular}{crrrrrrrrl}
\toprule
\textbf{Case} & \textbf{Day} & $\boldsymbol{P^0}$ & \textbf{Day$(P^0)$} & $\boldsymbol{\rho^0}$ & $\boldsymbol{P^u}$ & \textbf{Day$(P^u)$} & $\boldsymbol{\rho^u}$ & $\boldsymbol{\Delta\rho}$ & \textbf{Outcome} \\
\midrule
1 & 4 & 142 & 48 & 0.650 &   1 &  0 & 0.002 & 0.648 & Prevented \\
2 & 9 & 142 & 46 & 0.710 & 143 & 73 & 0.690 & 0.020 & Delayed \\
3 & 5 & 151 & 59 & 0.740 &   3 &  5 & 0.006 & 0.734 & Prevented \\
4 & 5 & 111 & 43 & 0.714 &   1 &  0 & 0.002 & 0.712 & Prevented \\
5 & 5 & 208 & 45 & 0.784 & 157 & 53 & 0.756 & 0.028 & Reduced  \\
6 & 5 & 125 & 75 & 0.660 &   2 &  5 & 0.004 & 0.656 & Prevented \\
\bottomrule
\end{tabular}
\end{table}

\begin{figure}[!htbp]
\centering
\includegraphics[width=\linewidth]{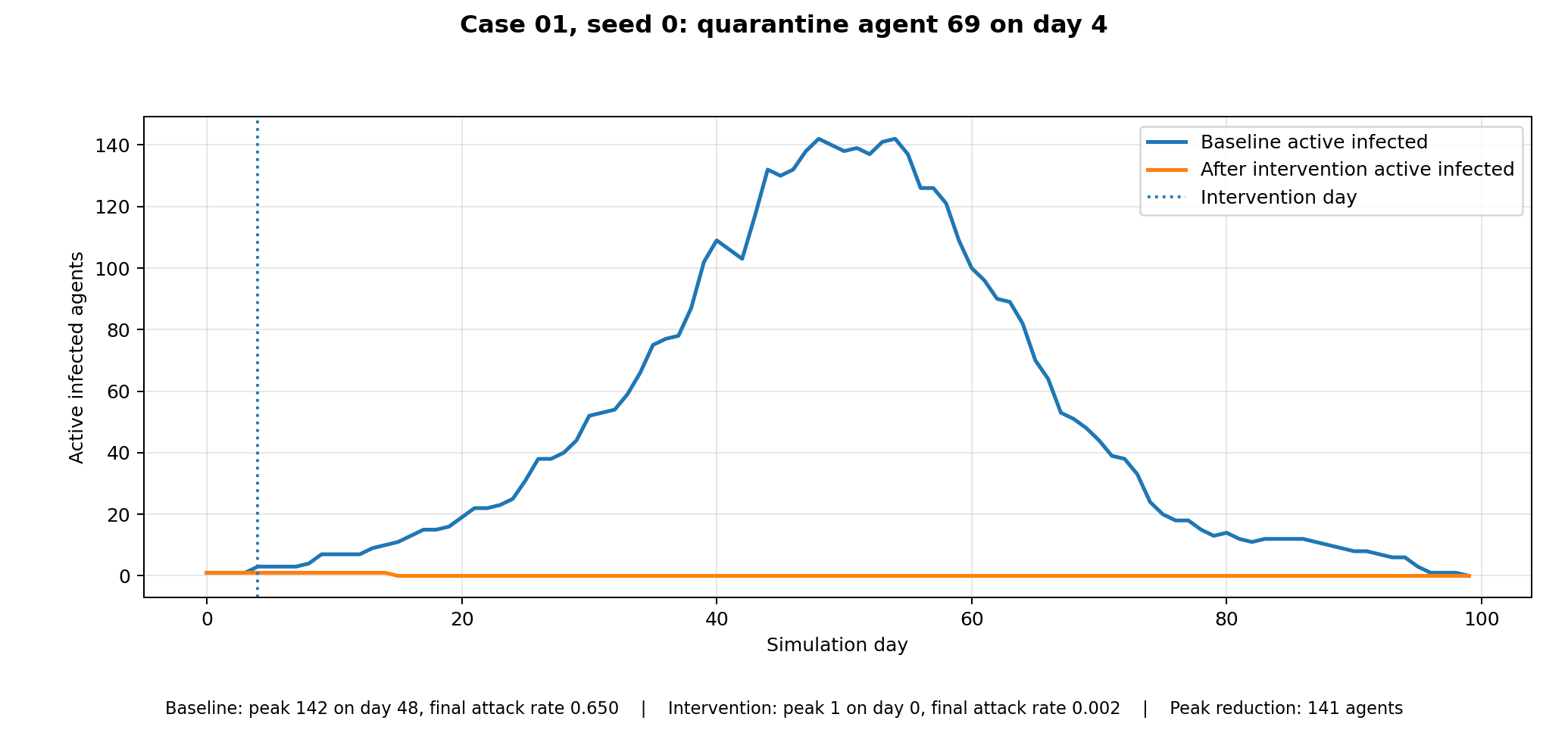}
\caption{Counterfactual intervention in Case~1. The baseline trajectory reaches a large active infection peak and a final attack rate of $0.650$. The one-day quarantine of the selected agent prevents sustained transmission, leaving a final attack rate of $0.002$.}
\label{fig:intervention-case-01}
\end{figure}

\begin{figure}[!htbp]
\centering
\includegraphics[width=\linewidth]{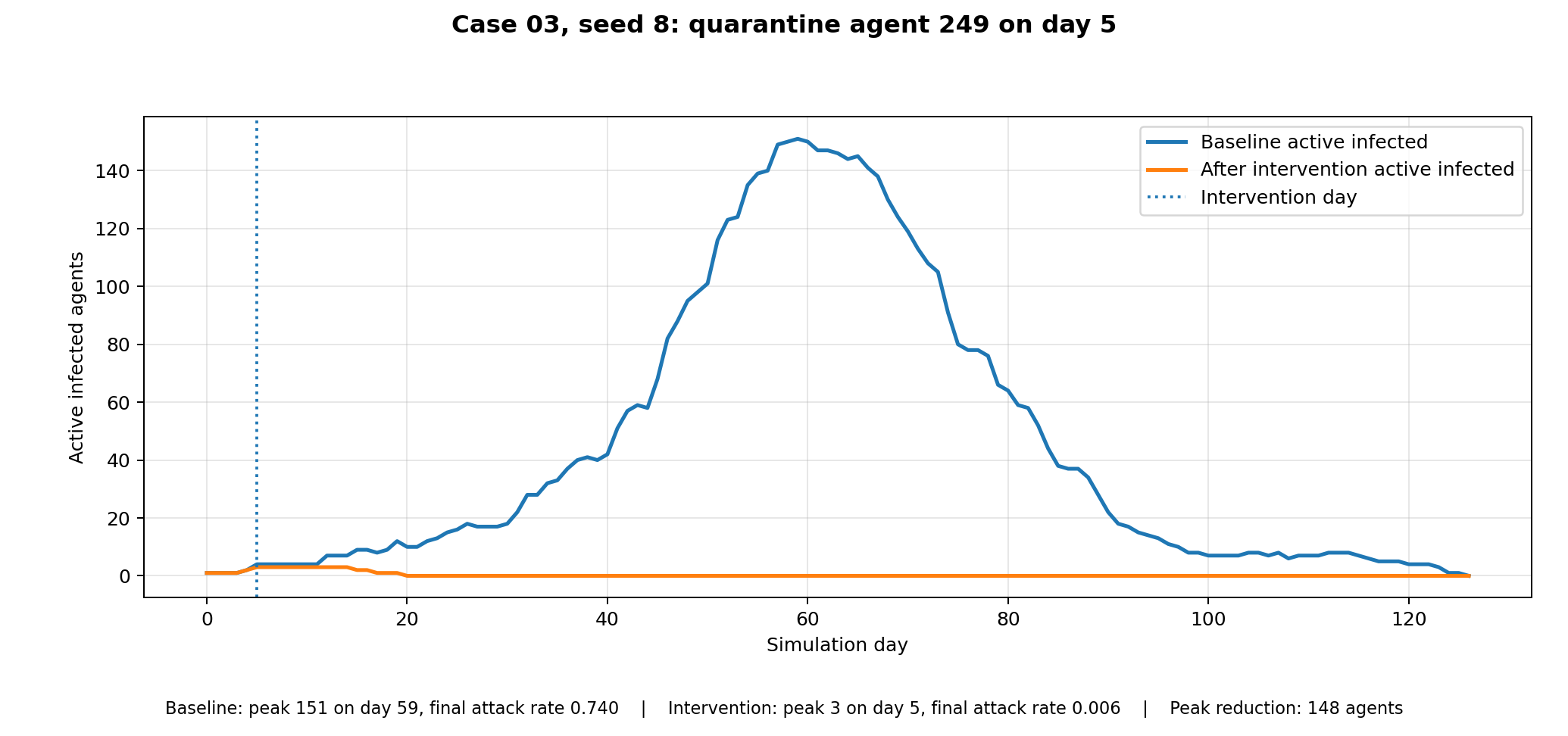}
\caption{Counterfactual intervention in Case~3. The baseline trajectory reaches a peak of 151 active infected agents and a final attack rate of $0.740$. The intervention suppresses the cascade, with a peak of 3 active infected agents and a final attack rate of $0.006$.}
\label{fig:intervention-case-03}
\end{figure}

\begin{figure}[!htbp]
\centering
\includegraphics[width=\linewidth]{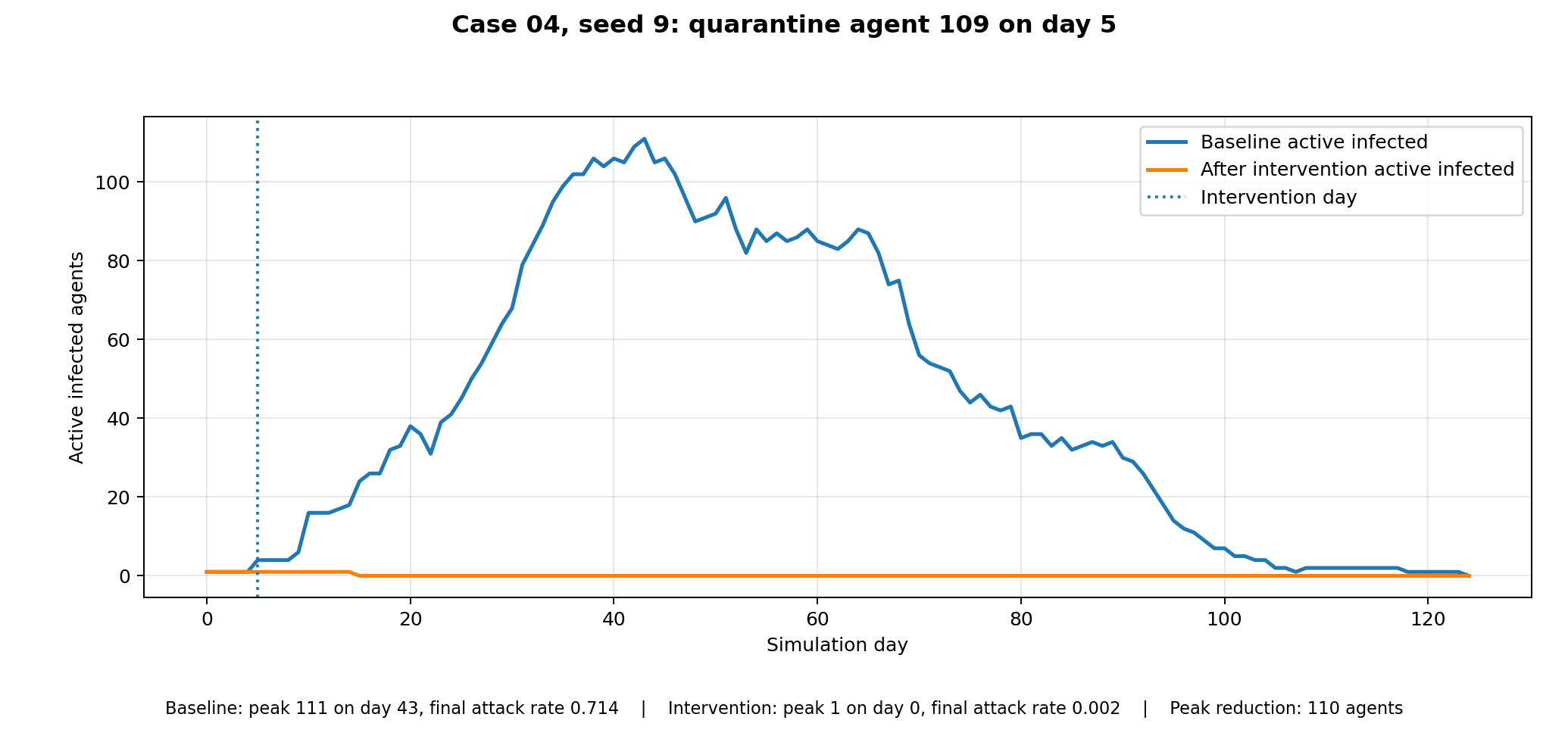}
\caption{Counterfactual intervention in Case~4. The baseline trajectory develops into a major outbreak, while the intervention trajectory remains close to extinction and reaches a final attack rate of $0.002$.}
\label{fig:intervention-case-04}
\end{figure}

\begin{figure}[!htbp]
\centering
\includegraphics[width=\linewidth]{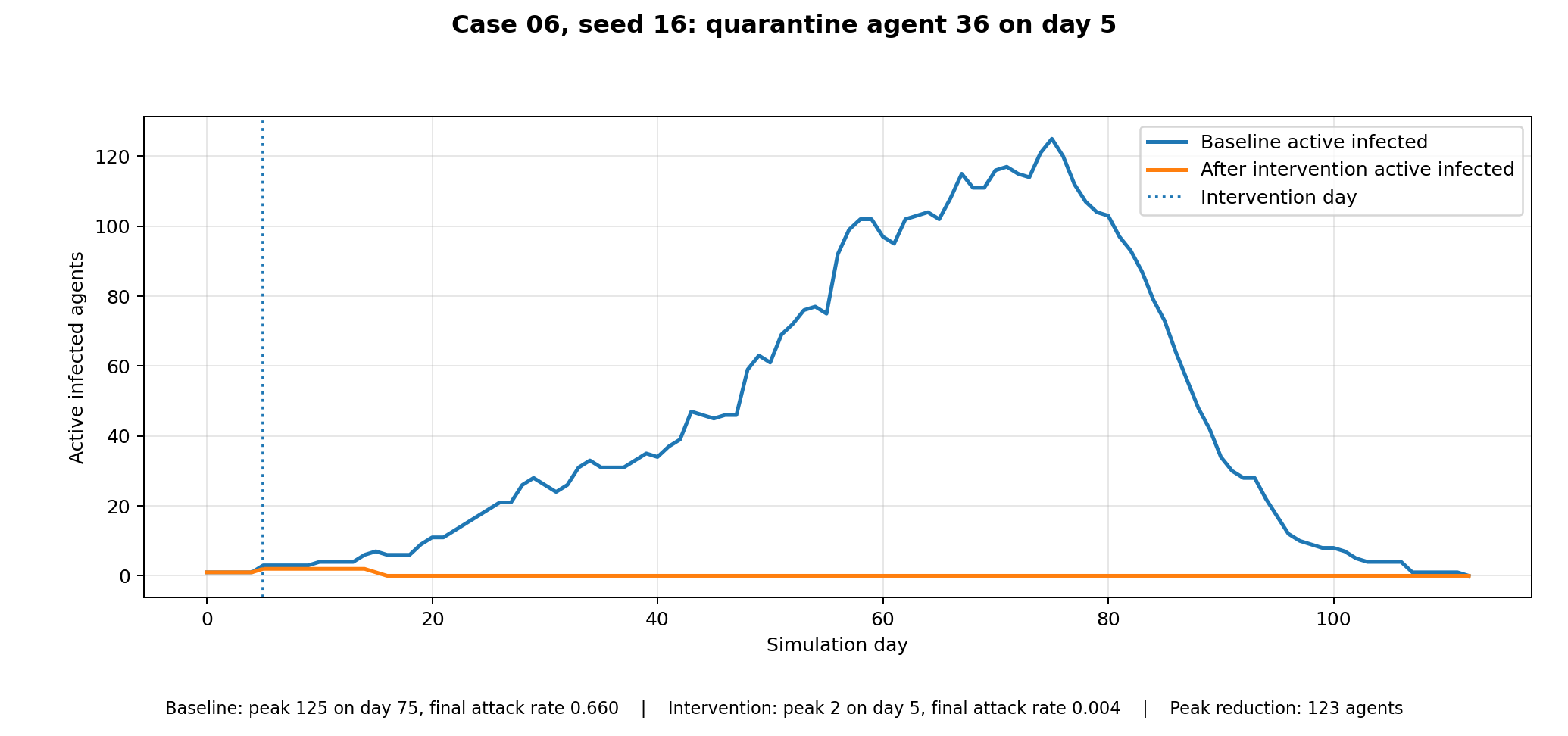}
\caption{Counterfactual intervention in Case~6. The baseline trajectory peaks late and reaches a final attack rate of $0.660$. The intervention prevents sustained transmission, with a peak of 2 active infected agents and a final attack rate of $0.004$.}
\label{fig:intervention-case-06}
\end{figure}

Figures~\ref{fig:intervention-case-01}, \ref{fig:intervention-case-03}, \ref{fig:intervention-case-04}, and~\ref{fig:intervention-case-06} show four examples in which the outbreak is prevented. In each figure, the blue curve is the baseline active infected count, the orange curve is the active infected count after intervention, and the vertical dotted line marks the intervention day. The baseline run begins with a small number of infections, enters a growth phase, and later reaches a substantial peak. The intervention run remains close to the initial infection level and then dies out. This behavior is consistent with a tipping-point interpretation. The intervention is small, but it occurs before the cascade becomes self-sustaining.

Figure~\ref{fig:intervention-case-01} shows the most direct form of prevention. The baseline trajectory grows slowly at first and then accelerates into a large wave. After the one-day quarantine, the active infected count remains near zero for the rest of the simulation. The large difference between the two curves indicates that the removed contact sequence lies upstream of a major downstream cascade.

Figure~\ref{fig:intervention-case-03} shows a similar prevention pattern. The baseline epidemic reaches a high active-infection peak, while the counterfactual trajectory remains close to extinction after the intervention. This case shows that the effect is not limited to a single trajectory shape; a minimal early restriction can redirect different outbreak trajectories when the interrupted contact sequence is sufficiently important.

Figure~\ref{fig:intervention-case-04} again shows a complete separation between the baseline and intervention outcomes. The baseline trajectory reaches a major outbreak regime, whereas the intervention trajectory stays at the extinction boundary. The result emphasizes that the relevant effect is not the size of the intervention itself, but its position in the causal contact sequence.

In Figure~\ref{fig:intervention-case-06} the baseline outbreak grows more slowly and peaks later than in some of the other examples, but the one-day intervention still prevents sustained spread. This case shows that preventability is not restricted to trajectories with an early visible explosion. A trajectory may remain modifiable while its early aggregate counts are still small.

\begin{figure}[!htbp]
\centering
\includegraphics[width=\linewidth]{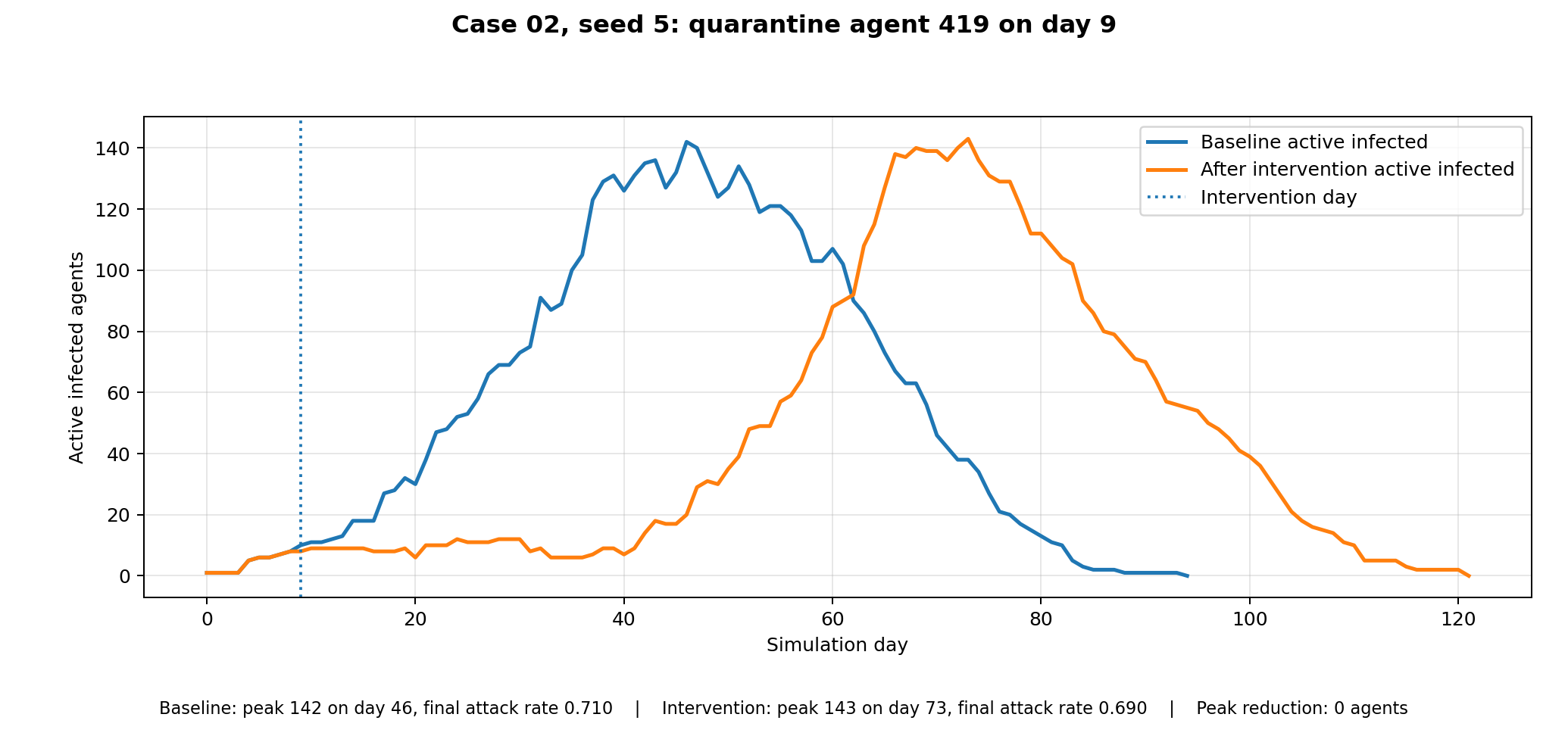}
\caption{Counterfactual intervention in Case~2. The one-day quarantine delays the epidemic peak, but the final attack rate remains above the major outbreak threshold.}
\label{fig:intervention-case-02}
\end{figure}

\begin{figure}[!htbp]
\centering
\includegraphics[width=\linewidth]{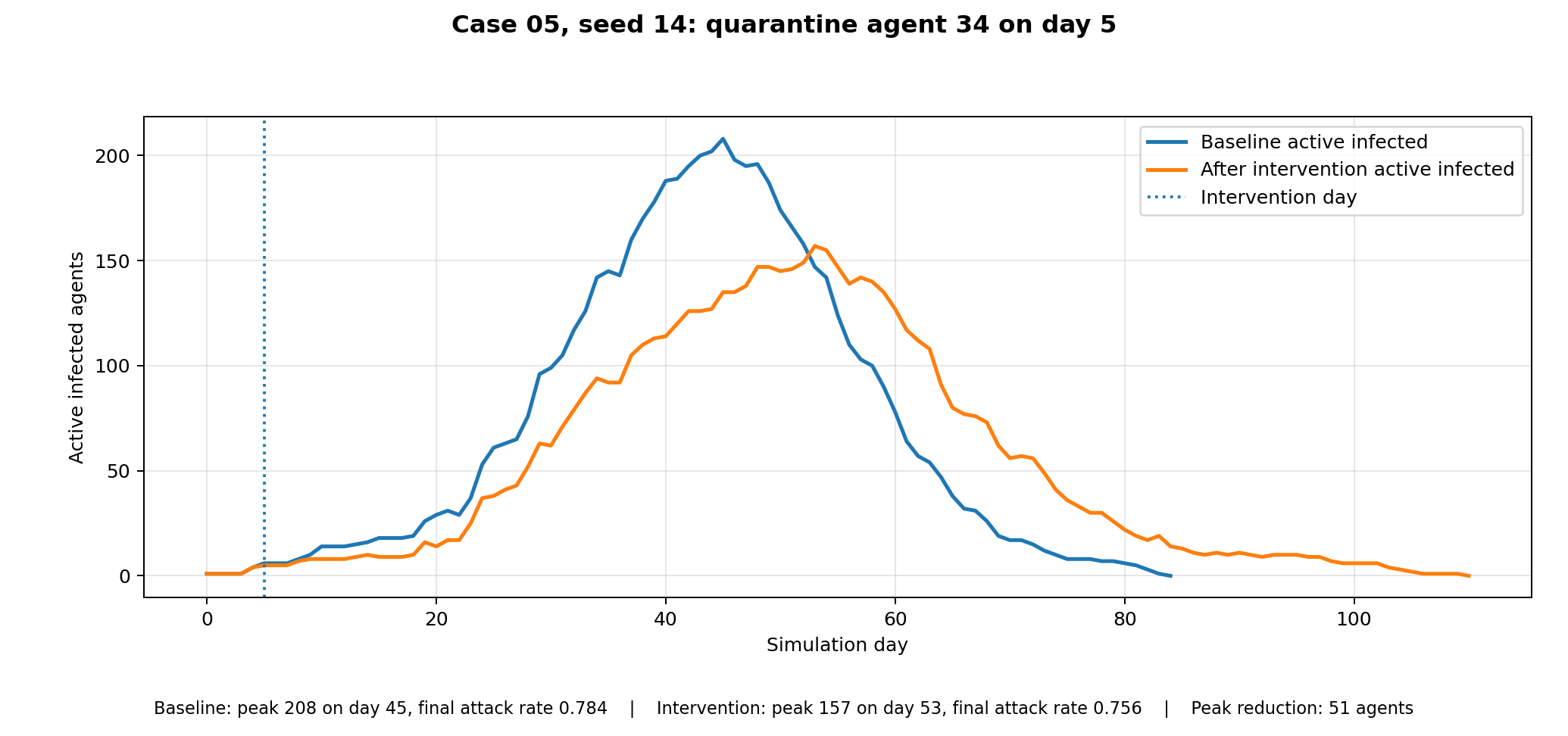}
\caption{Counterfactual intervention in Case~5. The intervention reduces and delays the active infection peak, but the final attack rate remains high and the trajectory remains a major outbreak.}
\label{fig:intervention-case-05}
\end{figure}

The two non-prevented cases are shown in Figures~\ref{fig:intervention-case-02} and~\ref{fig:intervention-case-05}. These examples show the limits of minimal intervention. A one-day quarantine can be decisive in some trajectories, but it does not guarantee prevention once other transmission paths are sufficient to sustain spread.

Figure~\ref{fig:intervention-case-02} illustrates a delay effect. The intervention shifts the epidemic peak from day~46 to day~73, but it does not reduce the final attack rate enough to change the outbreak label. The active infection peak is essentially unchanged. This means that the intervention alters the timing of spread, but does not remove enough downstream transmission to prevent the epidemic wave. In this case, alternative contact paths appear sufficient to preserve the outbreak.

Figure~\ref{fig:intervention-case-05} illustrates partial mitigation. The intervention reduces the peak active infected count from 208 to 157 and shifts the peak from day~45 to day~53. However, the final attack rate remains high. This indicates that the intervention weakens the epidemic wave but does not redirect the trajectory into the contained regime. Such cases are still meaningful because reducing the peak may matter for burden, even when the final outbreak classification does not change.

These results show that minimal quarantine interventions can have qualitatively different effects in the simulator. In several illustrative cases, a single one-day mobility restriction is sufficient to prevent sustained transmission. In other cases, the best evaluated intervention reduces or delays the epidemic but does not cancel it. They demonstrate that near-boundary epidemic trajectories can be highly sensitive to localized early perturbations, while trajectories with sufficient alternative transmission routes may remain robust to the same type of minimal intervention.

\FloatBarrier

\section{Conclusions}

The paper introduces a computational framework for studying early warning and minimal intervention in a nonlinear multi-agent epidemic simulation. The simulator provides a controlled environment in which local interactions between moving agents can generate qualitatively different global outcomes: rapid extinction, contained spread, or major outbreak. The central question is whether early aggregate trajectories contain enough information to identify the eventual regime and whether small, localized counterfactual changes can redirect trajectories near the outbreak boundary.

The results support this view. The boundary-focused simulations produce both contained and major outbreak trajectories under a narrow susceptibility sweep, which makes it possible to examine near-critical behavior rather than obvious low- or high-transmission cases. The Koopman model learns a compact latent representation of short epidemic histories and supports short horizon prediction through approximately linear latent dynamics. The latent projections are also correlated with the to final attack rate and outbreak class, which indicates that the learned representation captures meaningful information about the long-term regime.

The supervised early warning results further show that short observation windows can be highly informative. The random forest classifier achieves strong discrimination from five-day early windows, and the feature analysis indicates that Koopman-derived quantities contribute useful information alongside direct epidemic summaries such as active infections, incidence, and susceptible counts. Thus, the Koopman representation is not only a forecasting component, but also a useful source of features for downstream outbreak classification.

The counterfactual intervention experiments add a second layer to the analysis. They show that some outbreak trajectories remain highly modifiable early in their evolution. In several illustrative cases, quarantining one selected simulated agent for one selected day is enough to prevent sustained transmission and move the final attack rate below the major outbreak threshold. Other cases show weaker effects: the intervention reduces the peak burden without preventing the outbreak, or delays the outbreak while leaving the final size largely unchanged. The examples should not be interpreted as an estimate of a fixed intervention success rate; however, they show that the simulation can yield distinct dynamical regimes: in some trajectories, a small early perturbation interrupts a critical transmission chain, while in others, alternative contact paths are already sufficient to sustain spread.

The framework should be understood as a controlled computational study, not as a calibrated epidemiological decision system. The population, mobility rules, transmission mechanism, and intervention search are simplified by design. These assumptions make the counterfactual comparisons clear inside the simulator, but they do not directly specify a real-world control policy.

Future work could test the same framework under richer assumptions, including vaccination or contact network details. The intervention model could also be extended to multi-agent, multi-day, and cost-constrained policies. In addition, uncertainty quantification, comparisons with alternative latent dynamical models, and robustness tests across population sizes and mobility structures could clarify when early dynamical representations can reliably support intervention search in more realistic scenarios.

\bibliographystyle{plainnat}
\bibliography{references}

\end{document}